# HelixADMET: a robust and endpoint extensible ADMET system incorporating self-supervised knowledge transfer

Shanzhuo Zhang[1], Zhiyuan Yan[1], Yueyang Huang[1], Lihang Liu[1], Donglong He[1], Wei Wang[3], Xiaomin Fang[1,*], Xiaonan Zhang[1], Fan Wang[1,*], Hua Wu[2], Haifeng Wang[2]

[1] Baidu International Technology (Shenzhen) Co., Ltd., Shenzhen, China
[2] Baidu Inc., Beijing, China
[3] School of Computer Science and Technology, Harbin Institute of Technology (HIT), Shenzhen, China

* To whom correspondence should be addressed. Email: fangxiaomin01@baidu.com, wangfan04@baidu.com

**ABSTRACT**

Accurate ADMET (an abbreviation for "absorption, distribution, metabolism, excretion, and toxicity") predictions can efficiently screen out undesirable drug candidates in the early stage of drug discovery. In recent years, multiple comprehensive ADMET systems that adopt advanced machine learning models have been developed, providing services to estimate multiple endpoints. However, those ADMET systems usually suffer from weak extrapolation ability. First, due to the lack of labelled data for each endpoint, typical machine learning models perform frail for the molecules with unobserved scaffolds. Second, most systems only provide fixed built-in endpoints and cannot be customised to satisfy various research requirements. To this end, we develop a robust and endpoint extensible ADMET system, HelixADMET (H-ADMET). H-ADMET incorporates the concept of self-supervised learning to produce a robust pre-trained model. The model is then fine-tuned with a multi-task and multi-stage framework to transfer knowledge between ADMET endpoints, auxiliary tasks, and self-supervised tasks. Our results demonstrate that H-ADMET achieves an overall improvement of 4%, compared with existing ADMET systems on comparable endpoints. Additionally, the pre-trained model provided by H-ADMET can be fine-tuned to generate new and customised ADMET endpoints, meeting various demands of drug research and development requirements.

## INTRODUCTION

The modern drug discovery process is extremely time-consuming and expensive, and yet, only 10-20% of drug candidates can make it to the market (1, 2). The main reason for drug failure is usually the unsatisfactory bioavailability and toxicity (3) due to undesirable pharmacokinetic (PK) and pharmacodynamic (PD) properties. Therefore, accurately predicting PK/PD along with the absorption, distribution, metabolism, excretion, and toxicity (ADMET) can save both time and resources by filtering out unfavourable drug candidates at the early stage of drug development (4). Since the 1950s, medicinal chemists have been endeavouring to predict the aforementioned properties of candidate molecules (5). These works formed a set of specialised methodologies of quantitative structure-activity and structure-property relationships (QSAR and QSPR). Many of these methods are still highly applicable these days, such as AlogP and ClogP (6).

Traditionally, the prediction of molecular ADMET properties is considered the field of cheminformatics or medicinal chemistry (7–9). However, with the accumulation of massive data and our knowledge on the sophisticated interconnection between chemical and biological process, it is more and more recognized that a clear line shall not be drawn between the two fields (10). In fact, evaluation of molecular ADMET properties benefit from biological data, especially those reveal how molecules associate with clinical relevant endpoints (e.g., *in vivo* toxicity or side effects). This leads to a convergence of goals, tools and techniques of cheminformatics and bioinformatics (11) and creates needs to develop new models that fully utilise such combination of data from different domains.

Recent advances have shown great promise in applying machine learning techniques to mine the pattern and correlations from physicochemical and bioactivity data (12). Several comprehensive ADMET systems based on machine learning technology have been successfully developed to provide a variety of endpoint estimations, including admetSAR 1.0/2.0 (13, 14) (in 2012/2019), swissADME (15) (in 2017), vNN-ADMET (16) (in 2017), ADMETlab 1.0/2.0 (17, 18) (in 2018/2020), and FP-ADMET (19) (in 2021). Most of these systems favour traditional machine learning models, such as support vector machine (SVM), random forest (RF), and K-nearest neighbours (KNN), to estimate the molecular properties, i.e., endpoints. For example, FP-ADMET was constructed with more than 20 different molecular fingerprints as the input and RF as the machine learning model for over 50 ADMET endpoints. On the other hand, with the success of deep learning in various domains, deep learning models, especially graph neural networks (GNNs), have also been employed in ADMET prediction. admetSAR 2.0 applies a graph convolutional network (GCN) for building regression models, while ADMETlab 2.0 exploits an attention-based multi-task framework to train the model on multiple ADMET tasks simultaneously. Besides the application on ADMET tasks, GNNs also achieve state-of-the-art performance on large-scale molecular property prediction tasks maintained by the project of Open Graph Benchmark (20). Compared with traditional machine learning models, GNNs have shown their promise in ADMET prediction because they can encode molecules directly into graphs without the need of handcrafted molecular descriptors.

Although the existing ADMET prediction systems and methods have made a big step forward in technology, their extrapolation ability is still limited (21). First, there is a gap between the predictions of the commonly used machine learning models and the true value obtained from laboratory experiments, especially for molecules with unobserved scaffolds (core structures of molecules). The most important reason for the weak robustness of these models is the scarcity of labelled data due to costly and time-consuming laboratory evaluation. The scarcity of labels for ADMET endpoints makes sophisticated machine learning models easily overfit the limited training data and hard to generalise to the molecules with unobserved scaffolds. Second, even though the existing ADMET systems attempt to cover as many prediction properties/endpoints as possible, it is unfeasible in manpower for a system with only the built-in endpoints to fulfil all the prediction demands of drug researchers. Therefore, developing an ADMET system with endpoint extensibility, on which the researchers can build custom endpoints, is of great practical significance.

Recently, the technique of self-supervised learning (SSL) has proved its effectiveness in generating better molecular representation from large-scale unlabelled data and promoting the robustness of models. For example, Hu *et al.* (22) first proposed node-level and graph-level self-supervised tasks on GNNs for the generation of better molecular representation; in our previous work, we proposed multiple geometry-level SSL tasks to feed GNNs with molecular 3D spatial information (23). For the prediction tasks with only a small amount of labelled data, a more accurate prediction can be obtained by fine-tuning a model that has been pre-trained with large-scale self-supervised tasks. However, these works only focus on a few molecular properties, rather than developing a comprehensive system.

To this end, we develop a web-based comprehensive ADMET system, namely HelixADMET (H-ADMET, https://paddlehelix.baidu.com/app/drug/admet/train), to address the issue of weak extrapolation ability by SSL. We design a three-stage training framework incorporating the concept of self-supervised knowledge transfer. Self-supervised learning is utilised to learn general chemical knowledge from a large-scale unlabelled dataset, combined with a wide assortment of supervised tasks, to exploit the correlation among endpoints, auxiliary tasks, and self-supervised tasks. We evaluate the models thoroughly and demonstrate that H-ADMET substantially outperforms other ADMET systems with an improvement of 4%. H-ADMET can make excellent predictions even for the molecules with unobserved scaffolds. Moreover, in addition to the 52 built-in endpoints, H-ADMET also provides a pre-trained model, which can be fine-tuned with an external dataset to generate a highly accurate custom endpoint. Through fine-tuning, H-ADMET can be efficiently extended to cover theoretically unlimited endpoints and is expected to meet every specific drug research and development (R&D) requirement.

## SYSTEM DESCRIPTION AND METHODS

### ADMET profile

The ADMET profile, i.e., the endpoint selection of ADMET predictions, is constantly evolving. Overall, the ADMET prediction is a broader term, including but not limited to the topics about "PK/PD" and "drug-likeness prediction". In terms of drug development practice, all early predictions that are expected to reduce the risk of drug failure can be regarded as ADMET predictions. Therefore, compared to earlier PK/PD studies, a modern ADMET system contains a wider range of prediction endpoints. The ADMET endpoints provided by H-ADMET are shown in Figure 1. In particular, H-ADMET consists of 52 endpoints categorised into seven sections: a) physicochemical properties, b) medicinal chemistry, c) absorption, d) distribution, e) metabolism, f) excretion, and g) toxicity. We selected these endpoints by referring to multiple systems that are currently providing online ADMET prediction services, including SwissADME, ADMETlab 1.0/2.0, admetSAR 1.0/2.0 (13–15, 17, 18). Two factors are considered when selecting endpoints: 1) whether an endpoint is important in the early stage of drug development; 2) whether sufficient and high-quality data for an endpoint can be found, which determines the accuracy and robustness of the trained model. The details of data collections will be further discussed in the following section.

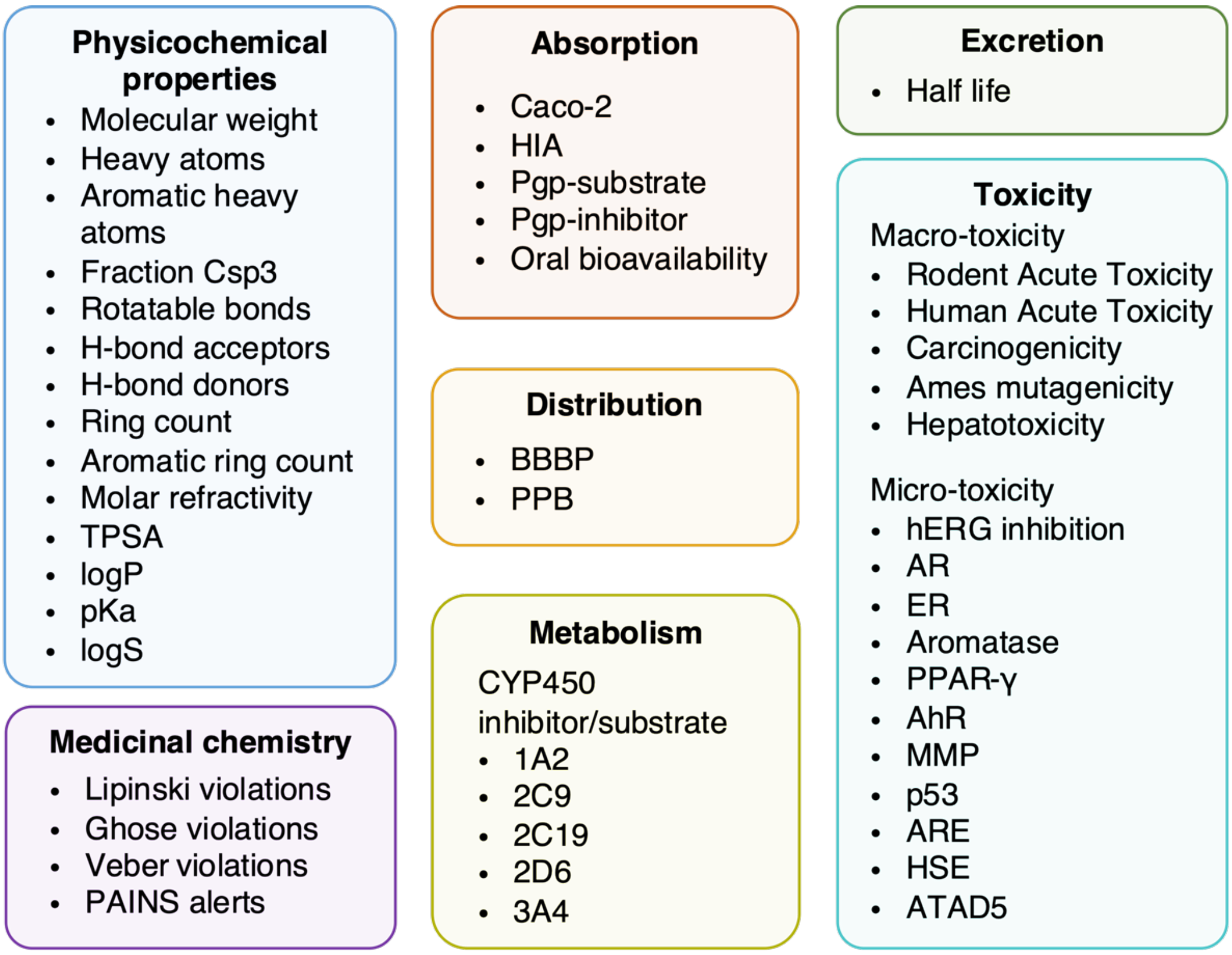

Figure 1. The ADMET profile provided by H-ADMET. Detailed descriptions of the endpoints are introduced in Supplementary Information.

There are some unique features in the design of toxicity endpoints of H-ADMET, as we notice that the *in vivo* toxicity of a compound can be seen on multiple physiological levels. For example, tetrodotoxin (TTX), one of the most potent natural toxins ever found, is a sodium channel blocker. It prevents neurons from firing action potentials by binding to voltage-gated sodium channels in nerve cell membranes and blocking the entering of sodium ions (which are important for the rising portion of an action potential) into the neuron. This stops signals from reaching the neurological system, and consequently muscles from contracting in response to nerve stimuli (24). The toxicity of TTX exhibits on multiple physiological levels: on the subcellular level, TTX inhibits sodium ion channels; on the cellular level, it prevents nerve and muscle cells from producing normal action potentials; on the organ level, this inhibition of cell behaviour can paralyse muscles, blood vessels, etc.; on the body level, it causes suffocation and death. Previous ADMET systems seldom consider the toxicity of compounds at different levels. In contrast, we divided the toxicity endpoints into macro-toxicity and micro-toxicity. The macro-toxicity includes the body and organ-level toxicity, while micro-toxicity includes cellular and sub-cellular level toxicity. We hope that this classification provides more insights into the toxicology of compounds to better assist drug development.

We adopted multiple approaches as models to make endpoint predictions. Some basic physicochemical endpoints were calculated with RDKit (version 2021_09_3) (25), such as molecular weight, number of heavy atoms, hydrogen-bond donor or acceptors, logP, etc. These basic endpoints constituted the input for calculating drug-likeness rules of medicinal chemistry, such as the Lipinski rule-of-five (26) and the Ghose and Veber rules (27, 28). The PAINS alert of medicinal chemistry defined some special molecular sub-structures to filter out false-positive compounds (29), which was also implemented by RDKit. More details about the algorithms of the above-mentioned endpoints can be seen in Supplementary Information. For the remaining endpoints, we utilised machine learning models that are trained on collected datasets and the details of which will be described in the following sections.

**Data collection and processing**

The collected data can be divided into two groups: unlabelled and labelled data. The large-scale unlabelled data are used in the self-supervised tasks, while the labelled data are the collection for ADMET, physicochemical endpoints, and other auxiliary bioactivity tasks used in the supervised tasks. The data collection is made freely accessible, please refer to the Supplementary Information for more details.

Unlabelled Data: The unlabelled data was extracted from the predefined drug-like subset in the ZINC15 database (30). This drug-like subset contains nearly 1 billion compounds, and we randomly selected 20 million compounds from the subset for self-supervised training. To the best of our knowledge, it is the first time that the self-supervised pre-training technology has been applied on open-accessed ADMET systems. The unlabelled dataset is also one of the largest ever used for ADMET prediction.

Labelled Data: The labelled data for ADMET and physicochemical endpoints and the auxiliary tasks was collected from various data sources, including DrugBank (31), ChEMBL (32), CPDB (33), PubChem assays (34), peer-reviewed scientific papers (35–41), and high-throughput screening projects such as Tox21 Data Challenge 2014 (42) (Tox21) and CYP450 (43). All compounds in the raw data are transformed into canonical SMILES strings by RDKit. We first combined the data extracted from various sources for each task, where special care was taken to ensure their experimental protocols, test species are completely identical. Then, we checked the molecules to remove invalid or duplicated ones (more details in Supplementary Information). We finally produced ADMET and physicochemical datasets with more than 50 thousand data entries and the auxiliary datasets with more than 500 thousand data entries, respectively. The data collection was highly diverse and sufficiently large enough to consider as a global dataset. Detailed statistics and scaffold analysis of our data collection are provided in Supplementary Table 3 and Supplementary Table 4.

**Training framework**

GNNs, specifically LiteGEM (44) and GINE+ (45), are the main machine learning models in H-ADMET (more details in Supplementary Information). The GNN models in H-ADMET are trained by a transfer learning framework, which will be introduced in detail. Meanwhile, we also adopt the traditional machine learning model RF as a complement in our system, whose performance has been verified in many other systems (14, 17, 19).

GNNs have been successfully applied to molecular property prediction in our previous work (23, 44). A molecule can be naturally represented as a graph, where the atoms and chemical bonds can be regarded as nodes and edges, respectively. Here, we design a three-stage training framework to train the GNNs to fully utilise the existing datasets and boost the performance of our model on the ADMET endpoints. The demonstration of the training framework is shown in Figure 2. In the first two stages, we leverage multi-task learning (MTL) to integrate supervised learning and SSL tasks together, transferring knowledge from one task to another. The parameters of the GNN backbone are shared by all the tasks, while the parameters of the head of each task are exclusive. Then, single-task fine-tuning is adopted for each endpoint to capture the task-specific information in the last stage, where each task is trained on its own GNN backbone and head.

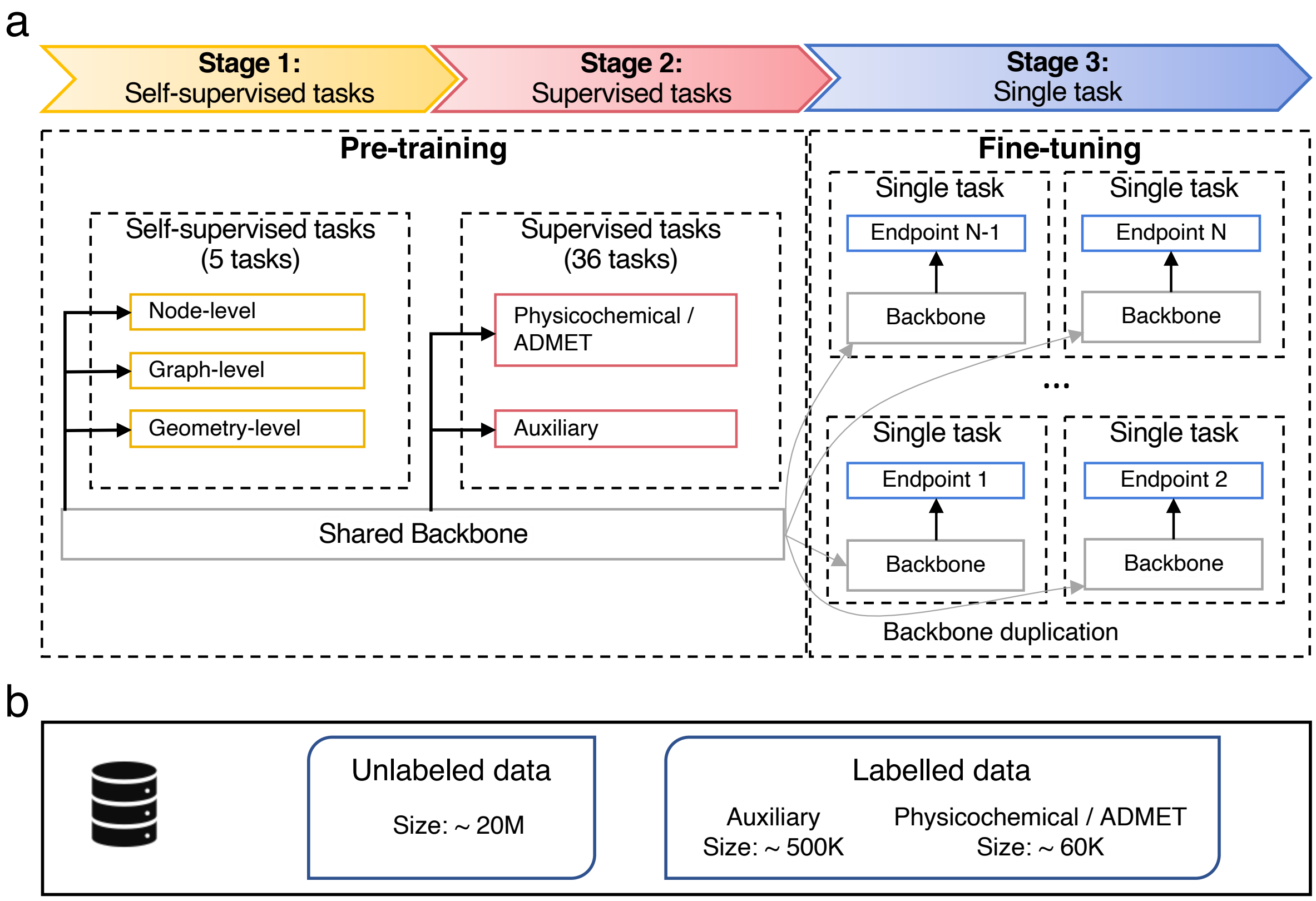

Figure 2. Demonstration of the training framework of H-ADMET. a) Detailed depiction of the training tasks and procedure in H-ADMET. b) The datasets and their scales in our data collection.

In the first stage, massive unlabelled molecules are utilised to pre-train the GNN with various SSL tasks to capture the general chemical knowledge. We adopt three levels of SSL tasks, including node/edge-level (22), geometry-level (23), and graph-level. 1) Node/edge-level task randomly masks a subgraph (local structure) of a compound from the model input, then trains the model to predict the masked subgraph to recover the whole graph. 2) For geometry-level tasks estimate, the model is trained to estimate the bond lengths and angles in a molecule. Thus, the model can learn in-depth physical laws from molecular 3D conformation. 3) The graph-level tasks let the model predict traditional molecular fingerprints. Molecular fingerprints encode information of molecular local structure (the ECFP fingerprint (46)) or functional groups (the MACCS fingerprint (47)) into vectors and may help the model acquire expert knowledge.

In the second stage, more than 40 supervised learning tasks, including physicochemical, ADMET, and auxiliary bioactivity tasks, are simultaneously trained. Note that the boundary of the first two stages is designed to be fuzzy, which means the second stage can overlap with the first stage, enabling the knowledge transfer from the self-supervised tasks to the supervised ones. Furthermore, except for the tasks of endpoints provided by the system (i.e., physicochemical and ADMET tasks), auxiliary tasks that estimate other bioactivities are also trained together in the second stage. The role of auxiliary tasks is to expand the scale of supervised training datasets. As shown in Figure 2a, the scale of auxiliary datasets is about ten times that of the physicochemical and ADMET dataset. By training on these more extensive domain data simultaneously, the performance and robustness of the model can be further improved (results further discussed in Supplementary Information).

In the third stage, we independently fine-tune each ADMET endpoint based on the model of the previous stage. By avoiding the interference of other endpoints, the model will focus on learning the unique information of each endpoint, and the parameters of the GNN backbone are no longer shared between different tasks. At the same time, the learning rate of the model at this stage will be reduced by ten times to ensure that the parameters of the model will not change drastically to retain the learned knowledge in the model to the greatest extent. Eventually, these fine-tuned independent models will be used for ADMET predictions and made available to users on our online system. In addition, this fine-tuning stage is also open to users, allowing users to fine-tune our pre-trained model by providing private data, thereby producing brand new and customised endpoints. More details of this function will be introduced in Results.

**Design of the web-based system**

We incorporated our models into a user-friendly, web-based system. Users can directly use our baseline model (trained by the three-stage framework) to generate predictions for 52 ADMET endpoints or choose to build their customised classification or regression models based on our pre-trained model. ADMET predictions based on the baseline model can be easily performed within two steps: model selection and query molecules input (Supplementary Figure 3 and Supplementary Figure 4). Then, the results will be arranged in a card view, and the "Advanced filters" panel on the result page allows the users to do a quick scan to exclude undesired compounds (Supplementary Figure 5 and Supplementary Figure 6). Alternatively, users may choose to build their customised models by uploading a training dataset containing molecules as SMILES strings and the corresponding ground truth labels. This uploaded dataset will be used to fine-tune the pre-trained GNN models in our system. After the fine-tuning is completed, the user can call this customised model at any time to make predictions. Since the pre-trained model can be fine-tuned on any external dataset, it provides users with better flexibility and extensibility and allows users to train exclusive models on any prediction endpoints of interest while maintaining the anonymity of the dataset. Altogether, we believe that H-ADMET is an easy-to-use and highly flexible system that generates comprehensive and accurate ADMET predictions.

## RESULTS

**Overall performance**

In the H-ADMET, a total of 36 prediction models were implemented (other endpoints were calculatable with RDKit), including 32 classification models and 4 regression models. For each endpoint, the best model is made publicly accessible on the system, which is chosen from LiteGEM, GINE+, and RF according to their performance on the corresponding test dataset. The detailed data splitting methods and training protocol are provided in Supplementary Information. The overall performance of H-ADMET and the other two well-known web-based ADMET systems, admetSAR 2.0 (14) and ADMETlab 2.0 (18), are shown in Table 1 and Table 2. The metrics used to evaluate classification and regression models are the area under the receiver operating characteristic curve (AUC) and the R-square ($R^2$), respectively. The AUCs on classification endpoints in H-ADMET range from 0.736 to 0.967 with an average value of 0.898. The $R^2$s of regression endpoints are all higher than 0.74.

Table 1. Comparison of the H-ADMET with other web-based systems on classification endpoints[a].

| Category | ADMET Endpoints | H-ADMET | admetSAR 2.0 (14) | ADMETlab 2.0 (18) |
|---|---|---|---|---|
| **Absorption** | Caco-2 permeability | **0.879** | 0.857 | - |
| | P-glycoprotein substrate | **0.891** | 0.865 | 0.840 |
| | P-glycoprotein inhibitor | **0.947** | 0.931 | 0.922 |
| | Oral bioavailability | 0.803 | - | **0.853** |
| **Distribution** | BBBP | **0.944** | 0.944 | 0.908 |
| **Metabolism** | CYP1A2 inhibitor | **0.948** | 0.883 | 0.928 |
| | CYP1A2 substrate | **0.949** | - | 0.737 |
| | CYP2C19 inhibitor | **0.939** | 0.871 | 0.913 |
| | CYP2C19 substrate | **0.945** | - | 0.758 |
| | CYP2C9 inhibitor | **0.934** | 0.858 | 0.919 |
| | CYP2C9 substrate | **0.944** | 0.625 | 0.725 |
| | CYP2D6 inhibitor | **0.905** | 0.840 | 0.892 |
| | CYP2D6 substrate | **0.956** | 0.772 | 0.847 |
| | CYP3A4 inhibitor | **0.930** | 0.848 | 0.921 |
| | CYP3A4 substrate | **0.967** | 0.695 | 0.776 |
| **Excretion** | Half life | 0.736 | - | - |
| **Macro-toxicity** | Carcinogenicity | 0.836 | **0.847** | 0.788 |
| | Hepatotoxicity | 0.808 | 0.719 | **0.814** |
| | Rodent Acute Toxicity | **0.717** | - | - |
| | Human Acute Toxicity | **0.873** | - | 0.853 |
| **Micro-toxicity** | Ames mutagenicity | 0.909 | **0.914** | 0.902 |
| | hERG inhibition | 0.909 | 0.811 | **0.943** |
| | AR | 0.858 | **0.886** | - |
| | ER | 0.848 | **0.880** | - |
| | PPAR-γ | 0.889 | 0.818 | **0.893** |
| | MMP | **0.951** | - | 0.927 |
| | p53 | **0.938** | - | 0.881 |
| | ARE | **0.917** | - | 0.863 |
| | HSE | 0.884 | - | **0.907** |
| | ATAD5 | **0.939** | - | 0.874 |
| | AhR | 0.922 | - | **0.943** |
| | Aromatase | **0.913** | 0.886 | 0.852 |
| | Average | **0.898** | - | - |
| | Comparison with admetSAR 2.0 [b] | **0.908** | 0.838 | - |
| | Comparison with ADMETlab 2.0 [b] | **0.914** | - | 0.866 |

[a] Endpoints provided by H-ADMET while not in admetSAR 2.0 or ADMETlab 2.0 will be denoted as “-”.

[b] Average AUC on the overlapping endpoints between H-ADMET and admetSAR 2.0 or ADMETlab 2.0

Table 2. Comparison of the H-ADMET with other web-based systems on regression endpoints.

| Category | ADMET Endpoints | H-ADMET | admetSAR 2.0 (14) | ADMETlab 2.0 (18) |
|---|---|---|---|---|
| Physicochemical property | Solubility | **0.877** | - | 0.850 |
| | pKa | **0.847** | - | - |
| Absorption | Human intestinal absorption | **0.786** | - | - |
| Distribution | PPB | **0.747** | 0.668 | 0.733 |
| | Average | **0.814** | - | - |
| | Comparison with admetSAR 2.0 | **0.747** | 0.668 | - |
| | Comparison with ADMETlab 2.0 | **0.812** | - | 0.792 |

Since the endpoints provided by different ADMET systems are usually not the same, in order to make a meaningful comparison with other systems, we calculate the average performance only on the overlapping endpoints between H-ADMET and each other system (Table 1 and Table 2). On overlapping classification endpoints, H-ADMET outperforms admetSAR 2.0 and ADMETlab 2.0 by 0.070 and 0.048 (AUC), respectively (Table 1). Meanwhile, the advantages of our system are approximately 0.079 and 0.020 ($R^2$) on overlapping regression endpoints, respectively (Table 2), which are also significant. Provided that the performance of the two baseline systems is already very high, it is quite challenging for our system to further achieve such a remarkable improvement. It is worth noting that a complete aligned comparison with admetSAR 2.0 and ADMETlab 2.0 is not possible since we could not replicate their databases, data pre-processing procedure, splitting methods, etc. A slightly better case is the comparison of 10 Tox21 endpoints (from PPAR-γ to Aromatase in Table 1), datasets of which are almost the same between our system and AMDETLab 2.0. The average AUC on these endpoints is 0.919 in H-ADMET, which is 0.026 higher than that of ADMETlab 2.0, indicating the superiority of our system. Although discrepancy exists, we believe that the results from different systems are still comparable to the extent that we only care about the performance of the model on unseen compounds, which has been certainly guaranteed when testing the model by all three systems.

### Ablation study on the three-stage training framework

We analyse the contribution of each stage in the training framework on LiteGEM. Both the random and scaffold split (details in Supplementary Information) are adopted to observe the performance of the ablation versions of the training framework. Compared with the random split, the scaffold split is more challenging since the scaffolds of the compounds in the test set will not appear in the training set. We compare three ablation versions that exploit part of the training stages, including *Stage 3*, *Stage 1* + *3*, *Stage 2* + *3*, to the original version of the proposed three-stage training framework, i.e., *Stage 1* + *2* + *3*. The average AUC values of the classification tasks under the random split and scaffold split are shown in Table 3. The detailed average AUC values on the classification tasks with the scaffold split are shown in Table 4.

Table 3. Overall contributions of each stage in the training framework on LiteGEM and classification tasks.

| Data Splitting Method | Stage 3 | Stage 1+3 | Stage 2+3 | Stage 1+2+3 |
|---|---|---|---|---|
| Random split | 0.850 | 0.855 | 0.882 | **0.887** |
| Scaffold split | 0.767 | 0.784 | 0.803 | **0.817** |

Table 4. Contribution of each stage in the training frameworks under the scaffold split on LiteGEM and classification tasks.

| Category | ADMET Models | Stage 3 | Stage 1+3 | Stage 2+3 | Stage 1+2+3 |
|---|---|---|---|---|---|
| Absorption [a] | Caco-2 permeability | 0.863 (0.009) | 0.870 (0.016) | 0.912 (0.014) | **0.957 (0.013)** |
| | P-glycoprotein inhibitor | 0.905 (0.006) | 0.909 (0.009) | 0.929 (0.006) | **0.932 (0.005)** |
| | Oral bioavailability | 0.704 (0.018) | 0.709 (0.024) | 0.739 (0.021) | **0.752 (0.019)** |
| Distribution | BBBP | 0.661 (0.019) | 0.681 (0.016) | 0.687 (0.015) | **0.695 (0.004)** |
| Metabolism | CYP1A2 inhibitor | 0.893 (0.005) | 0.900 (0.007) | 0.913 (0.002) | **0.916 (0.003)** |
| | CYP1A2 substrate | 0.760 (0.018) | 0.753 (0.009) | 0.764 (0.013) | **0.779 (0.023)** |
| | CYP2C19 inhibitor | 0.837 (0.009) | 0.859 (0.012) | 0.885 (0.004) | **0.886 (0.004)** |
| | CYP2C19 substrate | 0.675 (0.022) | 0.721 (0.005) | 0.735 (0.032) | **0.770 (0.021)** |
| | CYP2C9 inhibitor | 0.833 (0.007) | 0.866 (0.007) | 0.882 (0.007) | **0.896 (0.004)** |
| | CYP2C9 substrate | 0.729 (0.032) | 0.727 (0.018) | 0.746 (0.016) | **0.760 (0.018)** |
| | CYP2D6 inhibitor | 0.868 (0.011) | 0.876 (0.005) | 0.885 (0.005) | **0.895 (0.004)** |
| | CYP2D6 substrate | 0.717 (0.021) | 0.745 (0.017) | 0.747 (0.010) | **0.761 (0.016)** |
| | CYP3A4 inhibitor | 0.874 (0.004) | 0.888 (0.004) | 0.902 (0.003) | **0.911 (0.008)** |
| | CYP3A4 substrate | 0.670 (0.019) | 0.683 (0.028) | 0.690 (0.029) | **0.708 (0.018)** |
| Excretion | Half life | 0.718 (0.020) | 0.721 (0.019) | 0.730 (0.021) | **0.741 (0.018)** |
| Macro-toxicity | Carcinogenicity | 0.687 (0.017) | 0.706 (0.014) | **0.732 (0.026)** | 0.726 (0.016) |
| | Hepatotoxicity | 0.735 (0.034) | 0.765 (0.016) | 0.771 (0.015) | **0.780 (0.022)** |
| | Rodent Acute Toxicity | 0.642 (0.130) | 0.681 (0.062) | 0.723 (0.170) | **0.808 (0.119)** |
| | Human Acute Toxicity | 0.640 (0.010) | 0.647 (0.013) | 0.658 (0.009) | **0.661 (0.008)** |
| Micro-toxicity | Ames mutagenicity | 0.814 (0.006) | 0.824 (0.007) | **0.826 (0.005)** | 0.825 (0.021) |
| | hERG inhibition | 0.741 (0.009) | 0.771 (0.005) | 0.778 (0.010) | **0.788 (0.014)** |
| | AR | 0.875 (0.056) | 0.880 (0.011) | 0.888 (0.025) | **0.905 (0.028)** |
| | ER | 0.784 (0.021) | 0.816 (0.008) | 0.822 (0.013) | **0.861 (0.016)** |
| | PPAR-γ | 0.763 (0.040) | 0.791 (0.035) | 0.815 (0.012) | **0.825 (0.013)** |
| | MMP | 0.842 (0.016) | 0.852 (0.011) | 0.884 (0.009) | **0.900 (0.005)** |
| | p53 | 0.740 (0.041) | 0.763 (0.007) | 0.804 (0.006) | **0.826 (0.004)** |
| | ARE | 0.735 (0.023) | 0.759 (0.013) | 0.814 (0.016) | **0.816 (0.003)** |
| | HSE | 0.771 (0.037) | 0.786 (0.011) | 0.807 (0.016) | **0.817 (0.011)** |
| | ATAD5 | 0.767 (0.040) | 0.758 (0.006) | **0.819 (0.014)** | 0.806 (0.009) |
| | AhR | 0.824 (0.005) | 0.830 (0.010) | 0.847 (0.010) | **0.850 (0.014)** |
| | Aromatase | 0.742 (0.043) | 0.757 (0.008) | 0.782 (0.008) | **0.789 (0.004)** |
| | Average | 0.767 | 0.784 | 0.803 | **0.817** |

[a] The endpoint of P-glycoprotein substrate is removed because, under scaffold split, only one class remained in the test dataset, thus AUC cannot be calculated.

*Effect of self-supervised knowledge transfer.* We utilise three kinds of SSL tasks in the first training stage in order to learn general chemical knowledge. *Stage 1 + 3* incorporates unlabelled data by self-supervised knowledge transfer to boost the performance, while *Stage 3* only uses the task-specific data. *Stage 1 + 3* achieves a relative improvement of 0.6% and 2.2% over *Stage 3* under the random and scaffold split, respectively (Table 3). Besides, as we can see from Table 4, under the scaffold split, *Stage 1 + 3* performs better than *Stage 3* in 29/31 endpoints. The experimental results indicate that *Stage 1* can effectively improve the model's generalisation ability by learning knowledge from the large-scale unlabelled molecules and achieve better results on not only the unobserved molecules, but also the molecules with unobserved scaffolds.

*Effect of supervised knowledge transfer.* We assume that by adopting MTL to simultaneously train multiple supervised tasks, a task that estimates an endpoint can learn from the data of the correlated tasks. The AUC values of *Stage 3* and *Stage 2 + 3* in Table 3 and Table 4 verify our assumption. *Stage 2 + 3* that leverages self-supervised knowledge transfer achieves 3.8% and 4.7% improvement over *Stage 3* under the random split and scaffold split, respectively. The physicochemical and ADMET tasks as well as the auxiliary tasks contribute to promoting the models' performance. Please refer to Supplementary Information for more analysis of the ablation studies.

*Relationship between the self-supervised and supervised knowledge transfer.* Both self-supervised and supervised knowledge tasks have positive effects on performance promotion. Comparing *Stage 2 + 3* and *Stage 1 + 2 + 3*, we found that the improvement of *Stage 1 + 2 + 3* over *Stage 2 + 3* is not significant under the random split (results shown in Supplementary Table 10 and Supplementary Table 11), which reveals that there is an overlap between the knowledge learned by the self-supervised tasks and supervised tasks. Moreover, under the scaffold split, *Stage 1 + 2 + 3* emerges its superiority over *Stage 2 + 3* since self-supervised tasks have advantages in modelling the molecules with unobserved scaffolds. Accurately evaluating the molecular properties of novel scaffolds is highly significant for the discovery of new drugs.

### Model fine-tuning and unlimited endpoints

We carry out a proof-of-concept study to demonstrate that the pre-trained model produced by our training framework can efficiently generalise to unlimited new endpoints. In this study, we fine-tuned our pre-trained model with a small dataset of a new toxicity endpoint that was currently not included in the H-ADMET. The selected endpoint, drug-induced liver injury (DILI), is a leading cause of clinical trial failure and drug withdrawal (48). However, predicting DILI through a typical machine learning approach achieves limited success due to the scarcity of available data (just above 1k data entries) (49). In this study, a new model produced by H-ADMET was compared to that of vNN-ADMET (16), which offered a similar service to build customised models for new endpoints. We trained our model and the KNN model in vNN-ADMET on the same dataset curated by Xu *et al.* (50) (more details in Supplementary Information). The AUCs of the models generated by H-ADMET and vNN-ADMET on the same test dataset were 0.748 and 0.653, respectively. H-ADMET outperformed vNN-ADMET significantly, indicating that, through fine-tuning on a small dataset, H-ADMET is flexible and robust enough to be extended to previously unknown endpoints and make highly accurate predictions.

## DISCUSSION

From a physiological point of view, the correlation between ADMET tasks should be intuitive. To prove this correlation, Li *et al.* (51) pointed out that there was a correlation between multiple datasets of CYP450 inhibitors, and the utilisation of MTL on these datasets achieved better performance than using single-task training (51), which was further proved by Shen *et al.* (52). Based on these works, we assumed that ADMET endpoints, auxiliary tasks, and SSL tasks might all have correlations to some extent and verified this assumption on the CYP450 and Tox21 datasets. As shown in Figure 3, there is a strong positive correlation among CYP450 inhibitor endpoints (consistent with previous work) and among CYP450 substrate endpoints. On the contrary, a moderate negative correlation is observed between CYP450 inhibitor endpoints and substrate endpoints. These results are in line with our perception, as CYP450 enzymes all share similar structures, a compound that is an inhibitor of a certain

CYP450 enzyme is more likely to inhibit other CYP450 enzymes; and it is unlikely for a compound to be an inhibitor and a substrate of CYP450 enzymes at the same time. Interestingly, we also found that strong correlations also appear in unexpected endpoint pairs. For example, the correlation coefficient between the endpoint of AhR and CYP1A2 inhibitor is as high as 0.41. This correlation is beyond our knowledge when we developed the system but has been previously observed in experiments (53). We believe that by combining MTL and SSL, it is very convenient and natural to mine this kind of correlation with our knowledge transfer framework.

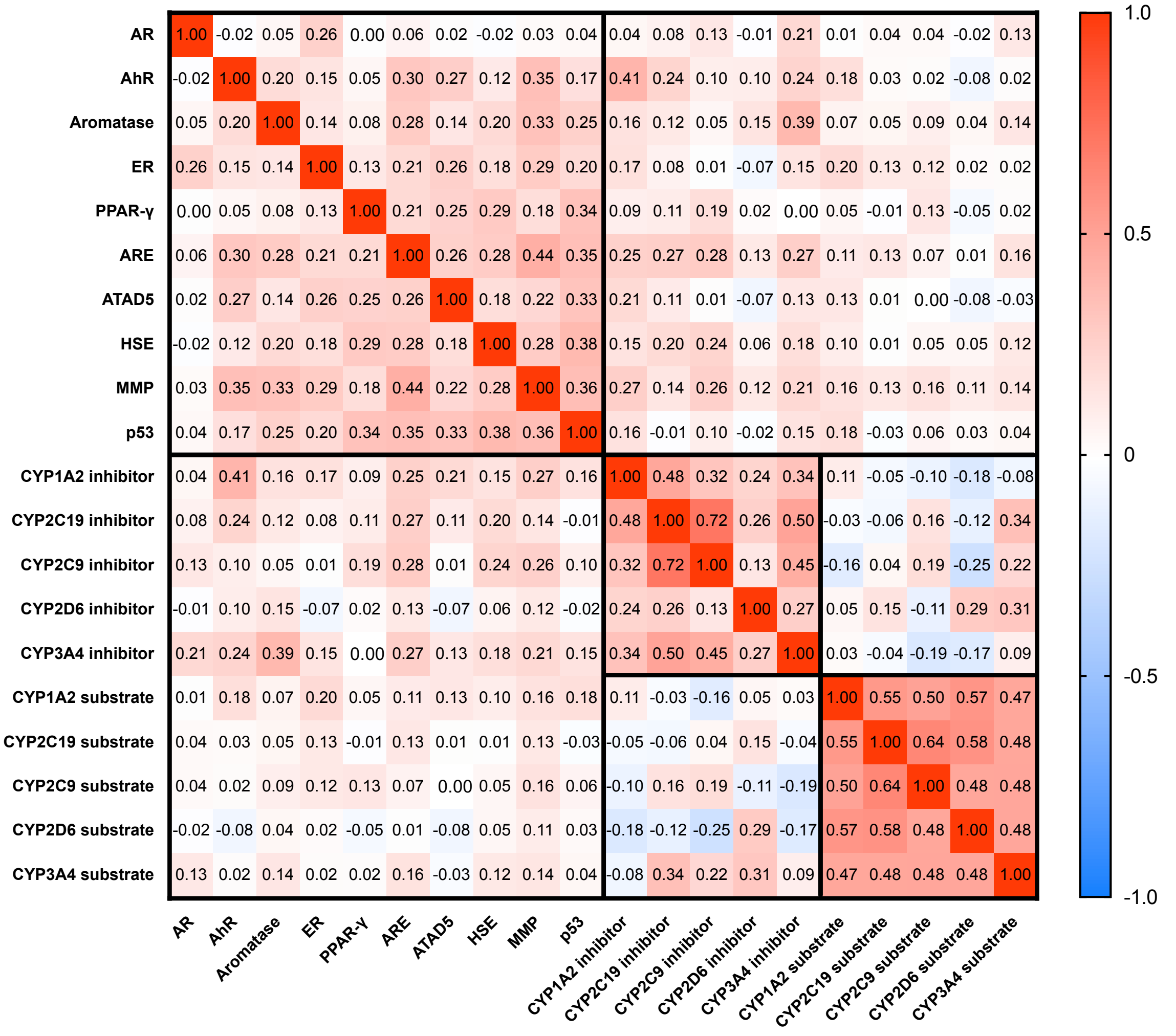

Figure 3. Correlations among the datasets of Tox21 and CYP450. The shared compounds were found for each pair of these tasks. Then the binary labels of these shared compounds were used to calculate the Pearson correlation coefficient. The abbreviations of Tox21 endpoints are defined in Supplementary Table 1.

At present, there are many web-based ADMET systems, such as swissADME (15), admetSAR 2.0 (14), and ADMETlab 2.0 (18), which can provide convenient online prediction services of ADMET properties. A) Unlike these systems, H-ADMET is designed with more focus on the flexibility of the system. In addition to 52 baseline endpoints, we allow users to fine-tune their models on our pretrained GNNs. Supplementary Figure 2 demonstrates that the new model fine-tuned on H-ADMET is effective and even outperforms the built-in endpoint in another system. B) In terms of model architectures, SwissADME mainly embeds traditional machine learning models, which are usually fast in prediction when applied to molecular design. admetSAR 2.0 uses a combination of traditional machine learning models and GCN for different endpoints. ADMETlab 2.0 connects an attention-based multi-task prediction network on a GCN backbone, which outperforms traditional tree-based models on most endpoints. We incorporated two state-of-the-art GNNs, LiteGEM and GINE+, and a traditional machine learning model, RF, in H-ADMET. The best model among the three will be chosen for each endpoint. C) A major difference between H-ADMET and other systems is the design of the multi-task and multi-stage training framework, which enables to transfer knowledge between ADMET, auxiliary, and self-supervised tasks. According to Supplementary Table 10, training stage 1 and 2 in our framework improved the average AUC of 0.037 on classification tasks, proving its effectiveness.

Besides precise prediction, user-friendliness also counts for an ADMET system in real-world drug discovery scenarios. For instance, when examining a large number of drug candidates, it is essential to quickly locate the ones of interest and keep an easily accessible record for all query molecules. H-ADMET is ideal for such projects given its advanced filters and task management module. Both are usually missing on other systems (Table 5). The filters on the result page save great effort by avoiding paying attention to unwanted results (Supplementary Figure 5). To help further accelerate the whole decision-making process, adverse results are also highlighted according to empirical rules (red marks shown in Supplementary Figure 6). In addition, all tasks run on the H-ADMET system are kept in the "Task Management" module upon account login. Users may rename, revisit, or download their prediction history at their convenience. This also applies to the new models trained with the user-provided dataset. That is to say, once users build a customised model for their property of interest, they can revisit the model without repeating the training process for their future prediction tasks. Overall, H-ADMET provides an accurate, flexible, well-organised, and user-friendly system for researchers to evaluate almost unlimited ADMET-related endpoints.

Table 5. Comparison of the key features of H-ADMET with other online systems.

| System | Customised models | Task management | Result filters | Indicators for negative signals | Capacity per batch | Number of Endpoints |
|---|---|---|---|---|---|---|
| H-ADMET | ✓ | ✓ | ✓ | ✓ | 1000 | 52* |
| FAF-Drugs4 (54) | ✗ | ✓ | ✓ | ✗ | - | 40 |
| admetSAR 2.0 (14) | ✗ | ✗ | ✗ | ✓ | 20 | 52 |
| vNN-ADMET (16) | ✓ | ✗ | ✗ | ✓ | - | 15* |
| ADMETlab 2.0 (18) | ✗ | ✗ | ✗ | ✓ | 500 | 88 |
| pkCSM (55) | ✗ | ✗ | ✗ | ✗ | 100 | 40 |
| SwissADME (15) | ✗ | ✗ | ✗ | ✗ | - | 44 |
| iDrug (56) | ✗ | ✓ | ✗ | ✗ | 100 | 69 |

✓ : the feature is supported; ✗ : the feature is not supported; "-": unspecified; *: Only the built-in endpoints are included, more endpoints can be evaluated by building customised models.

### Limitation and future work

To guide drug candidate prioritisation, there is an urgent need to provide a quantifiable drug-likeness score that summarises the overall ADMET profile. Traditionally, several experience-based rules proposed by Lipinski *et al.*, Veber *et al.*, Ghose *et al.,* and Gleeson *et al.* (26–28, 57) were widely used as filters to exclude the candidates that are unlikely to be orally available. Yet, these rules utilise stiff cut-offs and have limitations in their applications–oral drugs that break some of the rules are also increasingly reported (58, 59). To solve this problem, a more flexible indicator, Quantitative Estimate of Drug-likeness (QED), was proposed by Bickerton *et al.* to quantify drug-likeness (60), but QED does not take toxicity properties into consideration. Recently, the authors of admetSAR tried to define a scoring function named ADMET-score (61) based on 18 representative ADMET properties predicted by admetSAR. Although the results showed that ADMET-score was not markedly superior to QED in its ability to determine drug-likeness, it provided a new dimension in evaluating the overall ADMET performance, which is worth further exploration in the future.

Additionally, it is difficult to predict oral bioavailability and other endpoints of excretion and macro-toxicity because these endpoints are determined by multiple intermediate physiological processes. For example, the performance of our model on oral bioavailability is relatively poor (AUC = 0.803, see Table 1). A similar problem was also encountered in a recent work proposed by Fagerholm *et al.* (62). Even though our system can learn knowledge from multiple datasets, it still seems not satisfactory to train a fully end-to-end prediction model without the information of the intermediate processes. We may need to collect more datasets of the intermediate processes to enrich the knowledge and make further improvements.

## CONCLUSIONS

In this work, we present a highly robust and extensible ADMET system, namely HelixADMET, providing drug researchers with a convenient ADMET prediction service. A three-stage knowledge transfer training framework is proposed to leverage both MTL and SSL techniques, aiming to enhance the extrapolation ability of the system. Models in our system have been thoroughly evaluated with extensive ablation studies to demonstrate the effectiveness and superiority of our system. We also show that H-ADMET can make accurate predictions even on the molecules with unobserved scaffolds. Furthermore, we provide users with a model fine-tuning interface, allowing users to build new and customised endpoints. In this manner, H-ADMET is highly extensible and can cover unlimited endpoints theoretically. Altogether, we believe that H-ADMET is an easy-to-use and highly flexible system that can meet various demands of drug R&D requirements and help save both time and resources at the early stage of drug development.

## CONFLICT OF INTEREST

All authors declare that they have no conflicts of interest.

# Supplementary Information

**Supplementary Table 1.** Abbreviations of endpoints in HelixADMET.

| Endpoint category | Endpoint name | Abbreviation |
|---|---|---|
| **Physicochemical property** | Molecular weight | Mweight |
| **Physicochemical property** | Heavy atoms | nAtom |
| **Physicochemical property** | Aromatic heavy atoms | nAAtom |
| **Physicochemical property** | Fraction Csp3 | fCsp3 |
| **Physicochemical property** | Rotatable bonds | nRot |
| **Physicochemical property** | H-bond acceptors | nHA |
| **Physicochemical property** | H-bond donors | nHD |
| **Physicochemical property** | Ring count | nRing |
| **Physicochemical property** | Aromatic ring count | nARing |
| **Physicochemical property** | Molar refractivity | Refractivity |
| **Physicochemical property** | Topological polar surface area (TPSA) | TPSA |
| **Physicochemical property** | Lipid-water partition coefficient (log) | logP |
| **Physicochemical property** | Acid dissociation constant (pKa) | pKa |
| **Physicochemical property** | Water solubility (log) | logS |
| **Absorption (A)** | Caco-2 permeability | Caco-2 |
| **Absorption (A)** | Human intestinal absorption | HIA |
| **Absorption (A)** | P-glycoprotein substrate | Pgp-substrate |
| **Absorption (A)** | P-glycoprotein inhibitor | Pgp-inhibitor |
| **Absorption (A)** | Oral bioavailability | %F |
| **Distribution (D)** | Blood-brain-barrier penetration | BBBP |
| **Distribution (D)** | Plasma protein binding | PPB |
| **Metabolism (M)** | CYP1A2 inhibitor | - |
| **Metabolism (M)** | CYP1A2 substrate | - |
| **Metabolism (M)** | CYP2C19 inhibitor | - |
| **Metabolism (M)** | CYP2C19 substrate | - |
| **Metabolism (M)** | CYP2C9 inhibitor | - |
| **Metabolism (M)** | CYP2C9 substrate | - |
| **Metabolism (M)** | CYP2D6 inhibitor | - |
| **Metabolism (M)** | CYP2D6 substrate | - |
| **Metabolism (M)** | CYP3A4 inhibitor | - |
| **Metabolism (M)** | CYP3A4 substrate | - |
| **Excretion (E)** | Half life | $T_{1/2}$ |
| **Macro-toxicity (T)** | Acute oral toxicity on rodents | Rodent Acute Toxicity |
| **Macro-toxicity (T)** | Acute oral toxicity on human | Human Acute Toxicity |
| **Macro-toxicity (T)** | Carcinogenicity | - |
| **Macro-toxicity (T)** | Ames mutagenicity | - |
| **Macro-toxicity (T)** | Hepatotoxicity | - |
| **Micro-toxicity (T)** | hERG inhibition | - |
| **Micro-toxicity (T)** | Androgen (AR) receptor activation | AR |
| **Micro-toxicity (T)** | Estrogen (ER) receptor activation | ER |
| **Micro-toxicity (T)** | Aromatase (CYP19) inhibition | Aromatase |
| **Micro-toxicity (T)** | PPAR-γ activation | PPAR-γ |
| **Micro-toxicity (T)** | Aryl hydrocarbon receptor (AhR) activation | AhR |
| **Micro-toxicity (T)** | Mitochondrion membrane potential disturbance | MMP |
| **Micro-toxicity (T)** | p53 activation | p53 |
| **Micro-toxicity (T)** | Antioxidant response element activation | ARE |
| **Micro-toxicity (T)** | HSE activation | HSE |
| **Micro-toxicity (T)** | ATAD5 activation | ATAD5 |
| **Medicinal chemistry** | Lipinski violations | - |
| **Medicinal chemistry** | Ghose violations | - |
| **Medicinal chemistry** | Veber violations | - |
| **Medicinal chemistry** | PAINS alerts | PAINS |

# Description of all endpoints

## Physicochemical property

### 10. Molar refractivity

Molar refractivity is a measure of the total polarizability of a mole of substance. The method is described by Wildman *et al.*[1].

### 11. Topological polar surface area

TPSA is the surface sum over all polar atoms, primarily O and N, also including S and P atoms. The method is described by Ertl *et al.*[2].

**12. Lipid-water partition coefficient (ALogP)**

Partition coefficient is the ratio of a solute's concentration in n-octanol and its concentration in water, and ALogP is the calculated value for the 10-based logarithm of the ratio. It is a measure of the compound's hydrophilicity. The higher the value, the lower its hydrophilicity, which may cause poor absorption or permeation. The method is described by Wildman *et al.*[1].

**13. Acid dissociation constant (pKa)**

The pKa value evaluates the ability of an acid to release its hydrogen ion. A lower pKa value indicates a stronger acid, which dissociates more fully in water.

**Interpretation:**

| | | |
|---|---|---|
| <0 | : | Strong acid |
| 0-4 | : | Moderate acid |
| >4 | : | Weak acid |

**14. Water solubility (logS)**

logS is the 10-based logarithm of the solubility measured in mol/L unit. The larger the value, the higher the solubility. A compound with a logS < -6 usually exhibits poor solubility.

<u>**Absorption**</u>

**15. High Caco-2 permeability**

Caco-2 cells are widely used as an intestinal permeability model to predict the absorption of oral drugs[3]. The permeability of a compound is usually evaluated by Papp (Apparent permeability coefficient), which indicates the amount of compound that can be passively transported in a period of time, measured in cm/s. In general, a compound is considered to have a high Caco-2 permeability if it has a Papp>$8\times10^{-6}$ cm/s[4].

**Interpretation:**

The result is a real number in the range from 0 to 1. The closer the result is to 1, the higher the probability that the compound has a high Caco-2 permeability.

**16. HIA - Human intestinal absorption**

The percentage of the compound absorbed by the human intestine.

**Interpretation[5]:**

| | | |
|---|---|---|
| <30% | : | Poor absorption |
| 30%-79% | : | Moderate absorption |
| >=80% | : | Good absorption |

**17-18. P-glycoprotein substrate/inhibitor**

P-glycoprotein is a transporter protein located on the cell membrane. As an efflux pump, it pumps substance (including toxins and drugs) out of the cell. If a compound is a p-glycoprotein substrate, it is easily exported out of the cell, thereby a lower intracellular concentration. If a compound is a p-glycoprotein inhibitor, it prevents p- glycoprotein from transporting other compounds out of the cell, thereby increasing its intracellular concentration.

**Interpretation:**

The result is a real number in the range from 0 to 1. The closer the result is to 1, the higher the probability that the compound is a substrate/inhibitor of p-glycoprotein.

**19. Oral bioavailability (F)**

Oral bioavailability refers to the fraction of a drug that escapes gut-wall elimination, hepatic elimination and finally reaches the systemic circulation.

**Interpretation:**

The result is a real number in the range from 0 to 1. The closer the result is to 1, the higher the probability that the oral bioavailability of the compound is greater than 30%.

## Distribution

### 20. Blood-brain barrier permeability (BBBP)

BBBP determines whether a compound can enter the brain region. While getting through the blood-brain barrier is important for molecules that target specific region of the brain, it may also lead to potential side effects.

**Interpretation:**

The result is a real number in the range from 0 to 1. The closer the result is to 1, the higher the probability that the compound can pass the blood-brain barrier.

### 21. Plasma protein binding (PPB)

PPB refers to the fraction of a compound that attaches to the plasm proteins. A high PPB implies a higher distribution of the drug in the blood, rather than in other tissue. According to FDT (free drug theory), it is conventionally recognized that only unbound drugs may distribute into the tissue and act on its target(s).

**Interpretation:**

The result is a real number in the range from 0 to 100, the unit is %.

%PPB=100 * (DP-D)/DP, where DP is the concentration of drug-protein complex, and D is the concentration of unbound drug

A drug with a PPB value <85% usually exhibits good distribution in the tissue and would not lead to Drug-drug interaction (DDI) [1]

## Metabolism

### 22-31 Cytochromes P450 substrate/inhibitors

Cytochromes P450 (CYPs) are a superfamily of enzymes that oxidize other compounds in the body. They are the major enzymes that play a role in drug metabolism, participating in around 75% of the events. If a compound is the substrate of CYPs, it might be oxidized by the enzymes, result in inactive and/or toxic products. Meanwhile, activities of these compounds can be affected by other food/drugs containing CYP inhibitors. If a compound is a CYP inhibitor (AC50<10μM), it may increase the concentration of other drugs, result in Drug-drug interaction (DDI)-mediated toxicity.

**Interpretation:**

The result is a real number in the range from 0 to 1. The closer the value is to 1, the higher the probability that the compound is a substrate/inhibitor of the indicated CYP.

**Excretion**

**32. Half-life**

Half-life is the time (h) required for a compound to reach one-half its initial concentration in the body, it is an important measurement in pharmacokinetics. A short half-life (<3h) usually indicates a narrow therapeutic window, high toxicity, and more frequent dosing.

**Interpretation:**

The result is a real number in the range from 0 to 1. The closer the value is to 1, the higher the probability that the compound has a half-life greater than 3h.

**Toxicity**

**33. Acute oral toxicity on rodents (LD50)**

LD50 refers to the statistically derived dose that, when administered in an acute toxicity test on rodents, is expected to cause death in 50% of the treated animals in a given period. It is usually used to measure the toxicity of a compound. A compound with LD50<500mg/kg is considered moderate to extremely high toxic to rodents.

**Interpretation:**

The result is a real number in the range from 0 to 1. The closer the value is to 1, the higher the probability that the compound exhibits moderate to extremely high toxicity (<500mg/kg) in rodents.

**34. Acute oral toxicity on human (TDLo)**

TDLo is the lowest dose causing a toxic effect in acute oral toxicity test conducted in human. The lower the value, the higher the toxicity.

**Interpretation:**

The result is a real number in the range from 0 to 1. The closer the value is to 1, the higher the probability that the compound exhibits moderate to extremely high toxicity (<500mg/kg) in human.

**35. Carcinogenicity**

It indicates whether a compound may induce cancer in experimental animals.

**Interpretation:**

The result is a real number in the range from 0 to 1. The closer the value is to 1, the higher the probability that the compound will induce cancer.

**36. Ames mutagenicity**

The Ames test uses bacteria to test whether a given chemical can cause mutations in the DNA of the test organism. Ames mutagenicity is correlated to the carcinogenicity of a compound and can be used as a preliminary test for drug carcinogenicity.

**Interpretation:**

The result is a real number in the range from 0 to 1. The closer the value is to 1, the higher the probability that the compound will exhibit Ames mutagenicity.

**37. Hepatotoxicity**

It indicates whether a compound can cause liver damages including morphology changes or loss of function.

**Interpretation:**

The result is a real number in the range from 0 to 1. The closer the value is to 1, the higher the probability that the compound will cause hepatotoxicity.

**38. hERG inhibition**

hERG (the human Ether-à-go-go-Related Gene) is a gene (KCNH2) that codes for the potassium ion channel $K_v11.1$. The protein helps coordinate the heart's beating by mediating the repolarizing $I_{Kr}$ current in the cardiac action potential. hERG inhibition, which can lead to irregularity of the heartbeat and sudden death, should be avoided during drug discovery.

**Interpretation:**

The result is a real number in the range from 0 to 1. The closer the value is to 1, the higher the probability that the compound being an hERG inhibitor.

**39-43. Nuclear Receptor (NR) pathway-mediated toxicity**

NR pathways not only affect cell metabolism, growth, and proliferation, but also participate in intercellular signalling. As a result, compounds that act on NR pathways may trigger large scale adverse events. Influences on 5 NR pathways are used as indicators for the potential toxicity of a compound:

a) **Androgen receptor (AR) activation**
b) **Estrogen receptor (ER) activation**
c) **Aromatase (ARO) inhibition**
d) **Peroxisome proliferator-activated receptor gamma (PPAR-γ) activation**
e) **Aryl hydrocarbon receptor (AhR) activation**

The following table lists the consequences of disrupting each of these 5 NR pathways:

| | |
|---|---|
| **AR** | : Activation of which may trigger several diseases, such as prostate cancer. |
| **ER** | : Activation of which may trigger several diseases, such as breast cancer. |
| **ARO** | : Inhibition of which may lead to imbalance of androgen and estrogen. |
| **PPAR-γ** | : Activation of which may lead to weight gain, bone loss， congestive heart failure and etc. |
| **AhR** | :Activation of which may adversely affect a wide range of cell progression and function, and can lead to immunotoxicity. |

**Interpretation:**

The result is a real number in the range from 0 to 1. The closer the value is to 1, the higher the probability that the compound will affect the indicated pathway.

**44-48. Stress-induced cellular damage indicators**

External toxic compounds induce cellular stress after entering the cells, and may lead to cellular death. To assess the potential damage caused by a compound, 5 pathways are used as key indicators during toxicity screening.

- **a）Mitochondrion membrane potential (MMP) disturbance**
- **b）p53 activation**
- **c）Antioxidant response element (ARE) activation**
- **d）Heat shock factor response element (HSE) activation**
- **e）ATAD5 activation**

The following table lists the 5 pathways and their response under stress condition.

| | |
|---|---|
| **MMP** | :Disruption of which affects aerobic respiration in the cells. |
| **p53** | :Activated by stress conditions such as DNA damage. |
| **ARE** | :Activated by oxidative stress. |
| **HSE** | :Activated by heat shocks, leads to protein unfolding and affect cellular activities. |
| **ATAD5** | :Activated by DNA damage. |

**Interpretation:**

The result is a real number in the range from 0 to 1. The closer the value is to 1, the higher the probability that the compound will induce the indicated pathway.

**Drug likeness filter**

To guide the large-scale screening in the early stage of drug discovery, researchers proposed some 'Rules of thumb' based on their experience for selection of druglike compounds. It is generally believed that candidates meet these criteria are more likely to exhibit good pharmacokinetics properties, to display a higher bioavailability, and therefore more likely to be successful orally administrated drugs. Here we employ three classical filters to evaluate the drug-likeness of the compounds.

1. **Lipinski's rule of five/ Pfizer's rule of five/ rule of five (Ro5)**[6]
2. **Ghose Filter**[7]
3. **Veber Rule**[8]

The following table summarizes the main contents of the rules.

| | Ro5 | Ghose | Veber |
|---|---|---|---|
| **Molecular Weight (MW)** | <500 | 180 - 480 | |
| **octanol-water partition coefficient (clog P)** | <=5 | -0.4 - 5.6 | |
| **number of H-bond donors (NHD)** | <=5 | | |
| **number of H-bond acceptors (NHA)** | <=10 | | |
| **Molar refractivity (MR)** | | 40 - 130 | |
| **Total number of atoms** | | 20 - 70 | |
| **Rotatable bonds** | | | <=10 |
| **Topological polar surface area (TPSA)** | | | <=140 |

**Pan-assay interference compounds (PAINS) alert**

PAINS are chemical compounds that often give false positive results in high-throughput screens. These compounds tend to react non-specifically with numerous biological targets, lead to off-target effect and undesired side effects. Researchers summarize the functional groups frequently shared among PAINS as a structure filter, which can be used to flag candidate compounds. Here we employ the filter family A, B and C (in total 480 substructures) proposed in the literature[9]. One alert will be raised if a compound contains one of the substructures in the filter.

# Details of data collection and pre-processing

The dataset for each ADMET endpoint is usually extracted from various sources. In general, there are two steps in our data pre-processing protocol: 1) validation and deduplication, and 2) label processing. In the first step, we need to find the same compounds according to their canonical SMILES in the endpoint dataset and merge the duplicated ones according to their labels. For the classification endpoint, the final label of duplicated compounds will be determined with a voting method. The final label will be the one with more compounds. If the number of positive and negative labels are equal, the label of this compound is considered uncertain, and the compound will be discarded. For the regression endpoint, the final label is the average of all labels of the duplicated compounds. In the second step, we mainly check the labels of the regression endpoints, and make logarithmic transformations on labels to make the scale linear.

The processed dataset used by H-ADMET is made freely accessible at https://paddlehelix.bd.bcebos.com/HelixADMET_open_data/HelixADMET_open_data.tgz. There are ~2M unlabelled molecular data collected from ZINC[10], and labelled datasets from various literature and online databases (details shown in Supplementary Table 2). The dataset is shared under CC BY-NC-SA 4.0.

**Supplementary Table 2.** Data sources of the H-ADMET data collection.

| Dataset name | Data source |
|---|---|
| Caco-2 permeability | literature[4] |
| P-glycoprotein substrate | literature[11] |
| P-glycoprotein inhibitor | literature[12] |
| Oral bioavailability | literature[13] |
| Blood-brain-barrier permeability | literature[14] |
| CYP1A2 inhibitor | PubChem-AID1851 [a], SuperCYP[15] |
| CYP1A2 substrate | CypReact[16], SuperCYP[15] |
| CYP2C19 inhibitor | PubChem-AID1851 [a] |
| CYP2C19 substrate | CypReact[16], SuperCYP[15] |
| CYP2C9 inhibitor | PubChem-AID1851 [a] |
| CYP2C9 substrate | CypReact[16], SuperCYP[15] |
| CYP2D6 inhibitor | PubChem-AID1851 [a] |
| CYP2D6 substrate | CypReact[16], SuperCYP[15] |
| CYP3A4 inhibitor | PubChem-AID1851 [a] |
| CYP3A4 substrate | CypReact[16], SuperCYP[15] |
| Half life | DrugBank (collected in Nov, 2020)[17] |
| Acute oral toxicity on rodents (LD50) | ChemIDplus (collected in Nov, 2020) [b] |
| Acute oral toxicity on human (TDLo) | ChemIDplus (collected in Nov, 2020) [b] |
| Carcinogenicity | PubChem-AID1199, PubChem-AID1259411 [a] |
| Ames mutagenicity | literature[18] |
| Hepatotoxicity | literature[19] |
| Androgen receptor activation | Tox21 challenge14[20] |
| Estrogen receptor activation | Tox21 challenge14[20] |
| PPAR-γ activation | Tox21 challenge14[20] |
| Mitochondrion membrane potential disturbance | Tox21 challenge14[20] |
| p53 activation | Tox21 challenge14[20] |
| Antioxidant response element activation | Tox21 challenge14[20] |
| HSE activation | Tox21 challenge14[20] |
| ATAD5 activation | Tox21 challenge14[20] |
| Aryl hydrocarbon receptor activation | Tox21 challenge14[20] |
| Aromatase (CYP19) inhibition | Tox21 challenge14[20] |
| Water solubility (log) | literature[21,22] |
| Human intestinal absorption | literature[23] |
| Plasma protein binding | literature[24,25] |

[a] Data from PubChem can be found by searching with its AID on their website[26].

[b] Data was collect with the “Advanced search” service on ChemIDplus[27].

The statistics of our data collection for each endpoint is shown in **Supplementary Table 3**. Most of the datasets contain more than 1k compounds and the biggest one is the PubChem BioActivity (PCBA) dataset, which contains more than 400k compounds. The number of positive and negative entries in a dataset may not be the same because of the existence of empty values. A special treatment was given on the Tox21 dataset. Since the two assays of AR (PubChem AID 743040) and AR-LBD (PubChem AID 743053) in Tox21 dataset actually tested the same signalling pathway, we combined these two datasets as one toxicity endpoint. The same treatment was also applied on the ER and ER-LBD assays.

**Supplementary Table 3.** Statistics of the data collection in H-ADMET.

| Dataset name | Task type | Total entries | Positive entries | Negative entries |
|---|---|---|---|---|
| Fish & aquatic toxicity | Classification | 614 | 580 | 34 |
| Avian toxicity | Classification | 1434 | 316 | 1118 |
| HIV | Classification | 41127 | 1443 | 39684 |
| PubChem BioActivity dataset (PCBA) | Classification | 437929 | 15957 | 145067 |
| Caco-2 permeability | Classification | 528 | 235 | 293 |
| P-glycoprotein substrate | Classification | 571 | 312 | 259 |
| P-glycoprotein inhibitor | Classification | 1826 | 1198 | 628 |
| Oral bioavailability | Classification | 995 | 652 | 343 |
| Blood-brain-barrier permeability | Classification | 1965 | 1500 | 465 |
| CYP1A2 inhibitor | Classification | 12958 | 5989 | 6969 |
| CYP1A2 substrate | Classification | 1813 | 367 | 1446 |
| CYP2C19 inhibitor | Classification | 13047 | 5917 | 7130 |
| CYP2C19 substrate | Classification | 1795 | 296 | 1499 |
| CYP2C9 inhibitor | Classification | 12520 | 4175 | 8345 |
| CYP2C9 substrate | Classification | 1805 | 312 | 1493 |
| CYP2D6 inhibitor | Classification | 13555 | 2733 | 10822 |
| CYP2D6 substrate | Classification | 1844 | 394 | 1450 |
| CYP3A4 inhibitor | Classification | 12721 | 5286 | 7435 |
| CYP3A4 substrate | Classification | 1976 | 730 | 1246 |
| Half life | Classification | 893 | 597 | 296 |
| Acute oral toxicity on rodents (LD50) | Classification | 7646 | 3280 | 4366 |
| Acute oral toxicity on human (TDLo) | Classification | 289 | 255 | 34 |
| Carcinogenicity | Classification | 2151 | 1051 | 1100 |
| Ames mutagenicity | Classification | 7253 | 3951 | 3302 |
| Hepatotoxicity | Classification | 2169 | 1434 | 735 |
| hERG inhibition | Classification | 5356 | 2071 | 3285 |
| Androgen receptor activation | Classification | 7120 | 180 | 6940 |
| Estrogen receptor activation | Classification | 6680 | 406 | 6274 |
| PPAR-γ activation | Classification | 6450 | 186 | 6264 |
| Mitochondrion membrane potential disturbance | Classification | 5810 | 918 | 4892 |
| p53 activation | Classification | 6774 | 423 | 6351 |
| Antioxidant response element activation | Classification | 5832 | 942 | 4890 |
| HSE activation | Classification | 6467 | 372 | 6095 |
| ATAD5 activation | Classification | 7072 | 264 | 6808 |
| Aryl hydrocarbon receptor activation | Classification | 6549 | 768 | 5781 |
| Aromatase (CYP19) inhibition | Classification | 5821 | 300 | 5521 |
| Water solubility (log) | Regression | 2960 | - | - |
| Human intestinal absorption | Regression | 480 | - | - |
| Plasma protein binding | Regression | 1814 | - | - |
| Acid dissociation constant (pKa) | Regression | 6395 | - | - |

We adopted the Bemis-Murcko method implemented in RDKit to extract scaffolds from the compounds and to assess the chemical diversity of our data collection. The chirality of the scaffold was included and considered important when extracting the scaffolds. To illustrate the chemical variety of datasets, we estimated the ratio of molecules to scaffolds and the fraction of scaffolds with less than 5 molecules (results shown in Supplementary Table 4). The average number of molecules per scaffold varied from 1.3 to 3.5 in most datasets, with more than 95% of scaffolds containing no more than 5 molecules. The scaffold analysis revealed that the datasets contain a high amount of structural diversity, implying that models built with such datasets may better extrapolate to novel compounds.

**Supplementary Table 4.** Scaffold analysis of the data collection in H-ADMET.

| Dataset name | Unique scaffolds | Compounds / scaffolds | Ratio of scaffolds (<5 compounds) |
|---|---|---|---|
| Fish & aquatic toxicity | 104 | 5.904 | 0.923 |
| Avian toxicity | 271 | 5.292 | 0.878 |
| HIV | 19089 | 2.154 | 0.940 |
| PubChem Bioactivity dataset (PCBA) | 120084 | 3.647 | 0.825 |
| Caco-2 permeability | 314 | 1.682 | 0.946 |
| P-glycoprotein substrate | 415 | 1.376 | 0.981 |
| P-glycoprotein inhibitor | 1071 | 1.705 | 0.961 |
| Oral bioavailability | 640 | 1.555 | 0.967 |
| Blood-brain-barrier permeability | 1097 | 1.791 | 0.962 |
| CYP1A2 inhibitor | 7883 | 1.644 | 0.967 |
| CYP1A2 substrate | 883 | 2.053 | 0.963 |
| CYP2C19 inhibitor | 7898 | 1.652 | 0.966 |
| CYP2C19 substrate | 873 | 2.056 | 0.962 |
| CYP2C9 inhibitor | 7548 | 1.659 | 0.967 |
| CYP2C9 substrate | 881 | 2.049 | 0.963 |
| CYP2D6 inhibitor | 8186 | 1.656 | 0.966 |
| CYP2D6 substrate | 909 | 2.029 | 0.966 |
| CYP3A4 inhibitor | 7789 | 1.633 | 0.968 |
| CYP3A4 substrate | 982 | 2.012 | 0.960 |
| Half life | 598 | 1.493 | 0.980 |
| Acute oral toxicity on rodents (LD50) | 2035 | 3.757 | 0.937 |
| Acute oral toxicity on human (TDLo) | 142 | 2.035 | 0.972 |
| Carcinogenicity | 671 | 3.206 | 0.942 |
| Ames mutagenicity | 1749 | 4.147 | 0.899 |
| Hepatotoxicity | 1139 | 1.904 | 0.961 |
| hERG inhibition | 3066 | 1.747 | 0.952 |
| Androgen receptor activation | 2174 | 3.275 | 0.946 |
| Estrogen receptor activation | 2053 | 3.254 | 0.949 |
| PPAR-γ activation | 1864 | 3.460 | 0.946 |
| Mitochondrion membrane potential disturbance | 1719 | 3.380 | 0.947 |
| p53 activation | 2025 | 3.345 | 0.945 |
| Antioxidant response element activation | 1609 | 3.625 | 0.949 |
| HSE activation | 1856 | 3.484 | 0.947 |
| ATAD5 activation | 2112 | 3.348 | 0.943 |
| Aryl hydrocarbon receptor activation | 1985 | 3.299 | 0.945 |
| Aromatase (CYP19) inhibition | 1711 | 3.402 | 0.946 |
| Water solubility (log) | 690 | 4.290 | 0.899 |
| Human intestinal absorption | 335 | 1.433 | 0.985 |
| Plasma protein binding | 1040 | 1.744 | 0.967 |
| Acid dissociation constant (pKa) | 2018 | 3.169 | 0.909 |

# Details of model structures

Two GNN models were incorporated in H-ADMET. we use our previously proposed LiteGEM[28], which has obtained an excellent performance of KDD Cup 2021-PCQM4M-LSC, as the main model for the platform and use GINE+ as the baseline for the comparison. GINE+[29] is a state-of-the-art GNN baseline on the platform of open graph benchmark[30]. Also, we use the traditional machine learning algorithm RF as a kind of baseline in our system. RF is one of the most classical machine learning models and many literatures use RF as their baseline model because of its convenience for parameters tuning.

The model architecture of LiteGEM in H-ADMET is shown in **Supplementary Figure 1**. A molecule consists of atoms, and the neighboring atoms are connected by the chemical bonds, which can be naturally represented by a graph $G = (V, E)$, where $V$ is a node set and $E$ is an edge set. An atom in the molecule is regarded as a node $v \in V$ and a chemical bond in the molecule is regarded as an edge $(u, v) \in V$ connecting atoms $u$ and $v$. Basic GNN models can be seen as message passing neural networks[31], which are useful for predicting molecular properties. Following the definitions of the previous GNNs[32], the features of a node $v$ are represented by $x_v$ and the features of an edge $(u, v)$ are represented by $e_{uv}$. Taking node features, edge features and the graph structure as inputs, a GNN learns the representation vectors of the nodes and the entire graph, where the representation vector of a node $v$ is denoted by $h_v$ and the representation vector of the entire graph is denoted by $h_G$. A GNN iteratively updates a node's representation vector by aggregating the messages from the node's neighbors. Given a node $v$, its representation vector $h_v^{(k)}$ at the $k$-th iteration is formalized by

$$a_v^{(k)} = AGGREGATE^{(k)}\left(\left\{h_v^{(k-1)}, h_u^{(k-1)}, x_{uv} | u \in N(v)\right\}\right)$$

$$h_v^{(k)} = COMBINE^{(k)}\left(h_v^{(k-1)}, a_v^{(k)}\right)$$

where $\mathcal{N}(v)$ is the set of neighbors of node $v$, $AGGREGATE^{(k)}$ is the aggregation function for aggregating messages from a node's neighborhood, and $COMBINE^{(k)}$ is the update function for updating the node representation. We initialize $h_v^{(0)}$ by the feature vector of node $v$, i.e., $h_v^{(0)} = x_v$.

$READOUT$ function is introduced to integrate the nodes' representation vectors at the final iteration to gain the graph's representation vector $h_G$, which is formalized as

$$h_G = READOUT\left(h_v^{(k)} \middle| v \in V\right)$$

where $k$ is the number of iterations. In most cases, $READOUT$ is a permutation invariant pooling function, such as summation and maximization. The graph's representation vector $h_G$ can then be used for downstream task predictions.

Unlike convolutional neural networks, which can take advantage of stacking very deep layers, graph convolutional networks (GCNs) suffer from vanishing gradients, over-smoothing, and over-fitting issues when going deeper. To encode the whole molecular structures, motivated by DeeperGCN[33], we propose a message-passing strategy for graph convolution: LiteGEMConv, which is formalized as

$$m_{vu}^{(k)} = MLP\left(h_v^{(k-1)} \middle\| h_u^{(k-1)} \| e_{vu}\right)$$

$$a_v^{(k)} = SoftMax_{Agg_\beta}\left(\left\{m_{vu}^{(k)} \middle| u \in N(v)\right\}\right)$$

$$h_v^{(k)} = Linear\left(h_v^{(k-1)} + a_v^{(k)}\right)$$

where || denotes concatenation of vectors, $MLP$ denotes a 2-layer multi-layer perceptron (MLP) with SiLU as the activation function and $Linear$ denotes the Linear layer. $SoftMax_{Agg_\beta}$ function is used as our aggregation[33], which is defined as:

$$SoftMax_{Agg_\beta}(\cdot) = \sum_{u \in N(v)} \frac{exp(\beta m_{vu})}{\Sigma_{i \in N(v)}\, exp(\beta m_{vi})}$$

where $m_{vu} \in R^D$ is the given message set $\{m_{vu} | u \in N(v)\}$, and $\beta$ is the temperature controlling the smoothness of the distribution. Note that we also try to replace $SoftMax_{Agg_\beta}$ with a simple summation aggregation and observe no degradation in the performance, but we still keep it.

Overall, the LiteGEM consists of several LiteGEMConv layers with virtual node[31] representations added at each layer. Node features $h_v$ initialized as $x_v$ are updated by LiteGEMConv layer-by-layer, and the output is denoted as $h_v^{(k)}$. The graph level representation $h_G$ is obtained via mean pooling over the node representations of all nodes and the final output as the prediction of each endpoint is produced by another MLP as:

$$y_G = MLP_G(h_G)$$

Here $MLP_G$ is a 2- or 3-layer MLP with dropout, batch normalization and SiLU activation.

When the final output $y_G$ is obtained, we adopt different loss functions for different types of tasks to calculate the loss. Specifically, for the classification tasks, we use cross-entropy function $L_c$ to calculate the loss and for the regression task, we adopt L1 loss function $L_r$:

$$L_c = ylogy_G + (1-y)log(1-y_G)$$

$$L_r = \frac{\sum_{i=1}^{n} |y - y_G|}{n}$$

Where $y$ is the ground true label and $y_G$is the prediction of LiteGEM. It is worth noting that for each endpoint, we calculate the loss for each endpoint and backward propagate the gradient separately.

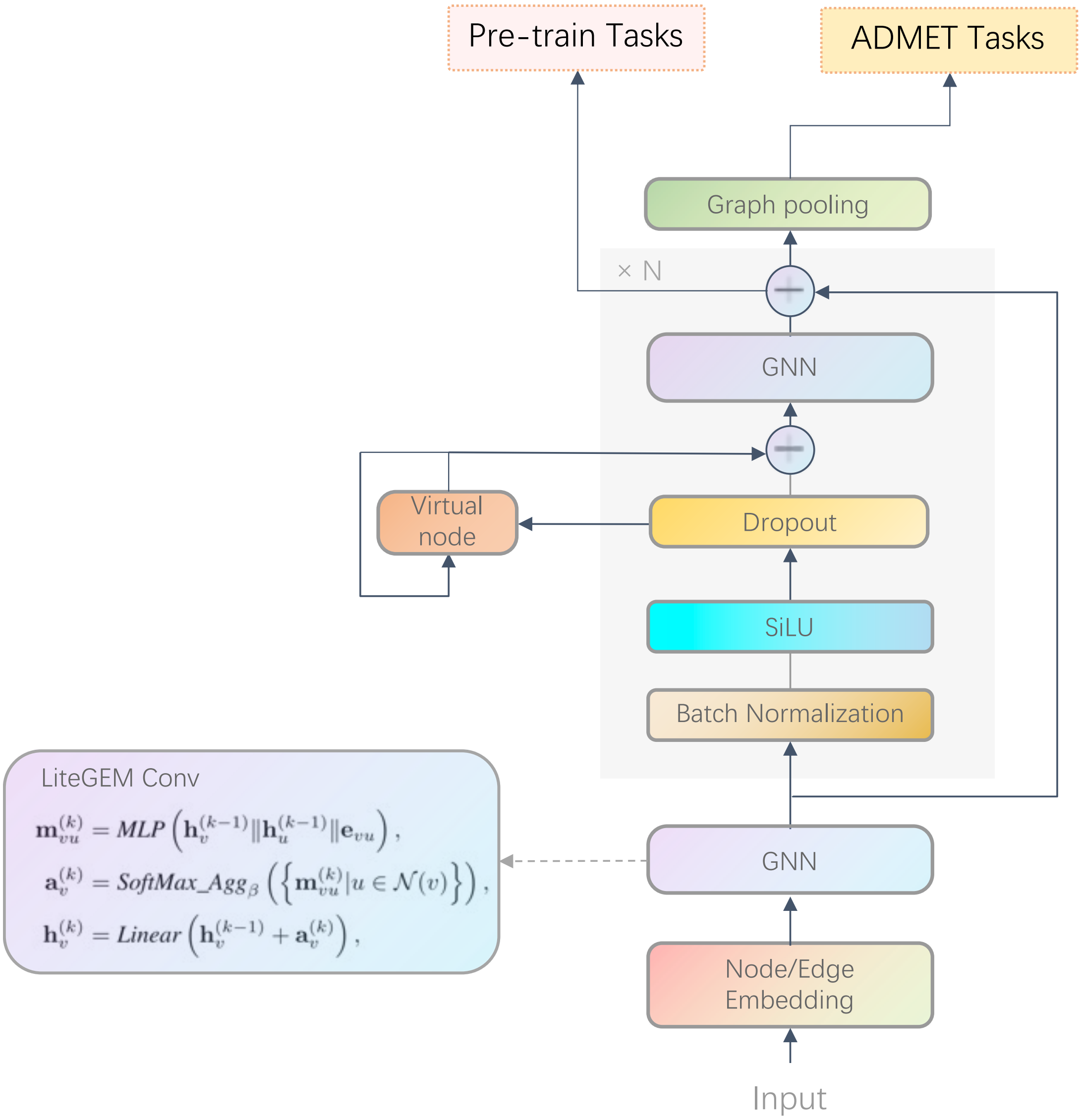

**Supplementary Figure 1.** The model structure of LiteGEM.

# Details of experiments

## Data splitting methods

In our experiment, we mainly used two data splitting methods—the random and scaffold split. When doing random split, each dataset is randomly split into a training set, validation set, and test set by the ratio of 8:1:1, with fixed random seeds ranging from 0 to 3 (4 independent runs for each hyper-parameter). The implementation of random splitting function and seed is provided by PaddlePaddle[34]. For the scaffold split, we have followed the method we adopted in GEM before[35]. Specifically, we first used the Bemis-Murcko method implemented in RDKit[36] to extract the scaffold of compounds, then put the compounds with the same scaffold into buckets and sort them in descending order according to the number of compounds in each bucket. Finally, a same splitting ratio of 8:1:1 was applied to divide the training set, validation set, and test set. A key difference between our scaffold splitting method and that in MoleculeNet[37] is that we considered the chirality of scaffolds, which made the test set more difficult for the model to predict. We found that the scaffold split was too difficult for models when trained on regression tasks. Therefore, only the random split was applied on regression tasks. To be noted, there is no randomness in the scaffold splitting method.

To demonstrate how the splitting methods will affect the difference of the valid/test sets comparing to the training set for each endpoint, we compared the similarity between these subsets of data for each endpoint under random or scaffold split, as shown in **Supplementary Table 5**. The similarity is defined as the average distance of all molecules in training set to their nearest molecules in the valid/test set. The distance is calculated by the Tanimoto similarity and the Morgan fingerprint with a length of 128. We can see that by applying the scaffold split, the similarities between subsets decreases significantly (about 0.05 – 0.15), which generates a more difficult benchmark and is helpful to improve the extrapolation ability of the model trained on it.

**Supplementary Table 5.** Similarity between training and valid/test set under random and scaffold split.

| Endpoints | Similarity (random split) | | Similarity (scaffold split) | |
|---|---|---|---|---|
| | train & valid | train & test | train & valid | train & test |
| Caco-2 permeability | 0.653(0.025) | 0.655(0.011) | 0.488 | 0.482 |
| P-glycoprotein substrate | 0.581(0.012) | 0.598(0.020) | - [a] | - [a] |
| P-glycoprotein inhibitor | 0.668(0.003) | 0.706(0.006) | 0.573 | 0.564 |
| Oral bioavailability | 0.567(0.018) | 0.578(0.010) | 0.535 | 0.563 |
| Blood-brain-barrier permeability | 0.640(0.015) | 0.651(0.005) | 0.588 | 0.546 |
| CYP1A2 inhibitor | 0.668(0.003) | 0.662(0.001) | 0.544 | 0.625 |
| CYP1A2 substrate | 0.613(0.015) | 0.633(0.017) | 0.561 | 0.505 |
| CYP2C19 inhibitor | 0.666(0.005) | 0.672(0.003) | 0.548 | 0.618 |
| CYP2C19 substrate | 0.605(0.016) | 0.630(0.017) | 0.550 | 0.491 |
| CYP2C9 inhibitor | 0.668(0.003) | 0.667(0.004) | 0.546 | 0.622 |
| CYP2C9 substrate | 0.620(0.016) | 0.617(0.021) | 0.561 | 0.509 |
| CYP2D6 inhibitor | 0.663(0.001) | 0.665(0.003) | 0.550 | 0.621 |
| CYP2D6 substrate | 0.623(0.013) | 0.624(0.015) | 0.568 | 0.509 |
| CYP3A4 inhibitor | 0.661(0.001) | 0.658(0.004) | 0.544 | 0.626 |
| CYP3A4 substrate | 0.638(0.010) | 0.656(0.007) | 0.576 | 0.587 |
| Half life | 0.544(0.012) | 0.542(0.029) | 0.501 | 0.521 |
| Carcinogenicity | 0.600(0.007) | 0.599(0.020) | 0.501 | 0.525 |
| Hepatotoxicity | 0.594(0.007) | 0.599(0.005) | 0.553 | 0.559 |
| Acute oral toxicity on rodents (LD50) | 0.639(0.001) | 0.638(0.002) | 0.519 | 0.536 |
| Acute oral toxicity on human (TDLo) | 0.511(0.029) | 0.520(0.015) | - [a] | - [a] |
| Ames mutagenicity | 0.700(0.004) | 0.703(0.008) | 0.571 | 0.576 |
| hERG toxicity | 0.705(0.007) | 0.703(0.006) | 0.668 | 0.662 |
| Androgen receptor activation | 0.648(0.007) | 0.643(0.006) | 0.543 | 0.542 |
| Estrogen receptor activation | 0.645(0.003) | 0.643(0.003) | 0.548 | 0.536 |
| PPAR-γ activation | 0.642(0.009) | 0.642(0.008) | 0.532 | 0.535 |
| Mitochondrion membrane potential disturbance | 0.635(0.004) | 0.635(0.001) | 0.535 | 0.524 |
| p53 activation | 0.640(0.005) | 0.645(0.006) | 0.545 | 0.539 |
| Antioxidant response element activation | 0.637(0.005) | 0.637(0.003) | 0.524 | 0.518 |
| HSE activation | 0.636(0.005) | 0.638(0.002) | 0.527 | 0.525 |
| ATAD5 activation | 0.655(0.003) | 0.646(0.002) | 0.539 | 0.541 |
| Aryl hydrocarbon receptor activation | 0.637(0.007) | 0.642(0.002) | 0.537 | 0.539 |
| Aromatase (CYP19) inhibition | 0.636(0.006) | 0.639(0.003) | 0.529 | 0.527 |
| Water solubility (log) | 0.644(0.004) | 0.640(0.011) | - [a] | - [a] |
| Acid dissociation constant (pKa) | 0.688(0.008) | 0.685(0.008) | - [a] | - [a] |
| Human intestinal absorption | 0.531(0.012) | 0.513(0.021) | - [a] | - [a] |
| Plasma protein binding | 0.597(0.014) | 0.595(0.006) | - [a] | - [a] |

[a] Scaffold split not applied on these endpoints.

# Model training

The LiteGEM and GINE+ were implemented in PaddlePaddle[34] and PGL[38]. The hyper-parameters searched when training the models are listed in **Supplementary Table 6**. Each group of hyper-parameters was trained four times with a fixed random seed (0, 1, 2, 3), and the mean and standard deviation of the metric of the best group were reported. The hyper-parameters searched for LiteGEM and GINE+ were aligned to make fair comparisons. However, due to the longer training time of GINE+, the hyper parameter searching range of GINE+ was halved compared with those of LiteGEM. All the GNN models were trained on the Nvidia Tesla A100. It took about 12h and 24h in training stage 1 for LiteGEM and GINE+ on 8 GPUs, respectively. In stage 2, LiteGEM took about 18-30 minutes each epoch to train, while GINE+ took about 30-40 minutes. Our results were generated with the following protocol: 1) epochs of stage 1 ranged from 0 to 10; 2) epochs of stage 2 ranged from 5 to 105; 3) epochs of stage 3 ranged from 105 to 110. The whole training time was about 2 days on a server with 8 GPUs. If the model was trained without stage 1 and stage 2, or in other words, trained in the single-task mode, a simple 100-epoch training protocol was adopted.

**Supplementary Table 6.** Hyper-params for all random/scaffold split runs.

| Hyper-parameters | Random Split | | Scaffold Split | |
|---|---|---|---|---|
| | LiteGEM | GINE+ | LiteGEM | GINE+ |
| **dropout rate** | 0.2 | 0.2 | 0.2 | 0.2 |
| **head layer number** | 2, 3 | 2, 3 | 2, 3 | 2, 3 |
| **learning rate** | 0.001, 0.005 | 0.001, 0.005 | 0.001, 0.005 | 0.001, 0.005 |
| **embedding dimension** | 128, 256 | 128 | 128, 32 | 128 |
| **GNN layer number** | 5, 8 | 5, 8 | 5, 8 | 5, 8 |

The RF model was implemented in scikit-learn[39]. The only hyper-parameter searched for RF was the number of estimators, which was set to 8, 32, or 128. RF was fitted to the same split datasets with the GNN models.

# Model fine-tuning and unlimited endpoints

We compared the performance of the fine-tuned model on our platform with other two ADMET systems (vNN-ADMET[40], ADMETlab 2.0[41]) based on the same dataset describing the drug-induced liver injury[42] (DILI). The dataset was randomly split into training, validation, and test sets (8/1/1) just as the splitting method used in previous experiments. The training set of 1065 compounds was then used to build new models on H-ADMET and vNN-ADMET platforms. And the three system were compared on a test set of 119 compounds. The same protocol was repeated for 4 times with random seed range from 0 to 3 and the averaged AUC and accuracy were shown in **Supplementary Figure 3**.

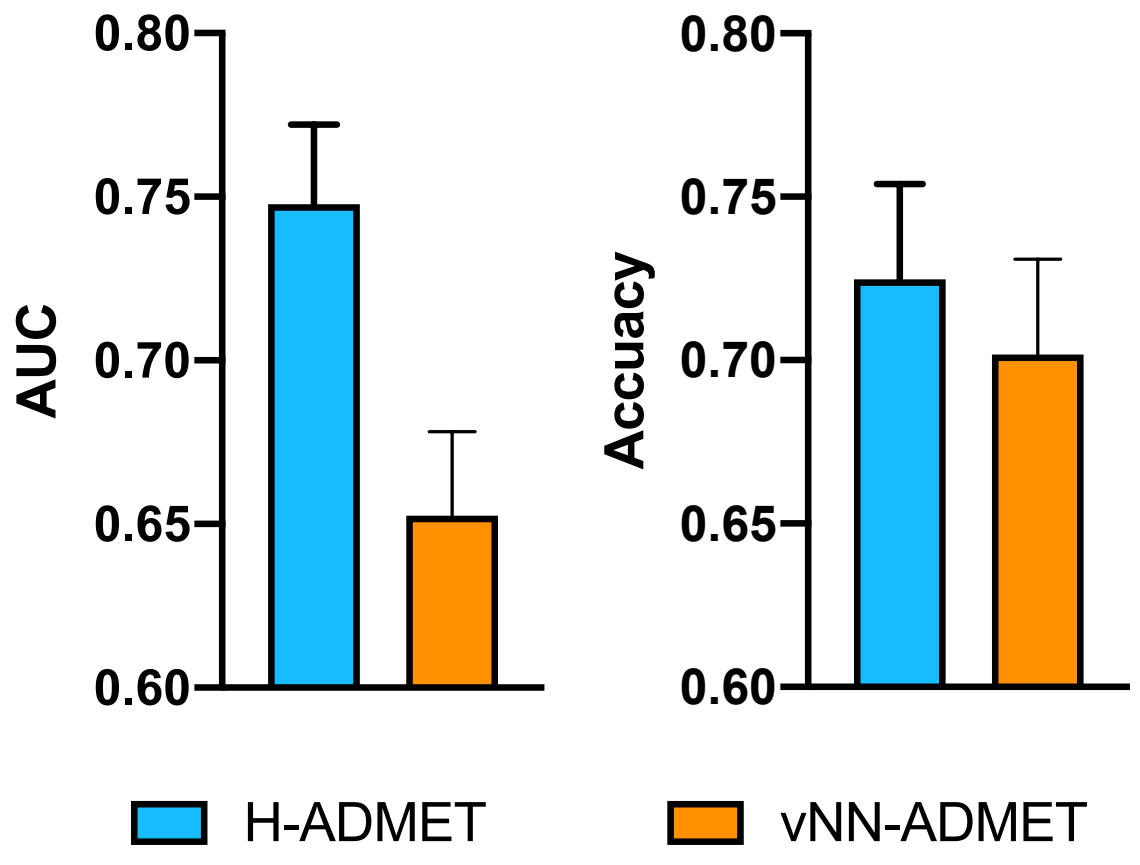

| | H-ADMET | vNN-ADMET |
|---|---|---|
| **AUC** | 0.748 (0.048) | 0.653 (0.051) |
| **Accuracy** | 0.725 (0.058) | 0.702 (0.058) |

**Supplementary Figure 2.** Comparison of a fine-tuned endpoint (DILI) between H-ADMET and vNN-ADMET. Performances of H-ADMET and vNN-ADMET on predicting DILI with a customised model, in terms of AUC and accuracy. Upper panel: All results are plotted as the average ± SD of four repeats. Lower panel: All results are summarised in the table, with SD in the bracket.

# Additional experiments

**Effect of the strategy of incorporating bioactivity datasets.** We also incorporated bioactivity datasets other than the ADMET tasks, such as PCBA[43] and HIV[44], enabling simultaneous multi-task learning to boost task diversity. The reason for using these datasets is mainly due to their large scale, while ADMET datasets are typically quite small, as shown in **Supplementary Table 3**. In addition, it enables the model to perform transfer learning, i.e., to learn domain knowledge for the prediction of ADMET endpoints. **Supplementary Table 7** provides more details than Table 4, in which we can see that, with scaffold split, the addition of bioactivity datasets improves both the model without pre-training and the model with pre-training, 1.0% and 0.7%, respectively. It is worth noting that we have only randomly selected a few large-scale bioactivity datasets as auxiliary tasks, however, we believe that if we can increase the diversity of the datasets further, the final results should be relatively better.

**Supplementary Table 7.** The contribution of bioactivity datasets as auxiliary tasks under the scaffold split.

| Category | ADMET Models | Stage 2+3 | Stage 1+2+3 | Stage 2+3 | Stage 1+2+3 |
|---|---|---|---|---|---|

| | | w/o Bio[a] | w/o Bio | | |
|---|---|---|---|---|---|
| Absorption | Caco-2 permeability | 0.939 (0.013) | 0.918 (0.032) | 0.912 (0.014) | **0.957 (0.013)** |
| | P-glycoprotein inhibitor | 0.922 (0.005) | 0.930 (0.003) | 0.929 (0.006) | **0.932 (0.005)** |
| | Oral bioavailability | 0.728 (0.020) | 0.737 (0.016) | 0.739 (0.021) | **0.752 (0.019)** |
| Distribution | BBBP | 0.699 (0.005) | **0.705 (0.009)** | 0.687 (0.015) | 0.695 (0.004) |
| Metabolism | CYP1A2 inhibitor | 0.905 (0.003) | 0.911 (0.003) | 0.913 (0.002) | **0.916 (0.003)** |
| | CYP1A2 substrate | 0.762 (0.006) | **0.782 (0.013)** | 0.764 (0.013) | 0.779 (0.023) |
| | CYP2C19 inhibitor | 0.864 (0.003) | 0.865 (0.004) | 0.885 (0.004) | **0.886 (0.004)** |
| | CYP2C19 substrate | 0.707 (0.029) | 0.763 (0.018) | 0.735 (0.032) | **0.770 (0.021)** |
| | CYP2C9 inhibitor | 0.860 (0.004) | 0.880 (0.007) | 0.882 (0.007) | **0.896 (0.004)** |
| | CYP2C9 substrate | **0.766 (0.017)** | 0.729 (0.016) | 0.746 (0.016) | 0.760 (0.018) |
| | CYP2D6 inhibitor | 0.869 (0.001) | 0.888 (0.004) | 0.885 (0.005) | **0.895 (0.004)** |
| | CYP2D6 substrate | 0.726 (0.004) | **0.765 (0.021)** | 0.747 (0.010) | 0.761 (0.016) |
| | CYP3A4 inhibitor | 0.888 (0.006) | 0.901 (0.008) | 0.902 (0.003) | **0.911 (0.008)** |
| | CYP3A4 substrate | 0.670 (0.046) | **0.738 (0.028)** | 0.690 (0.029) | 0.708 (0.018) |
| Excretion | Half life | 0.722 (0.016) | 0.729 (0.019) | 0.730 (0.021) | **0.741 (0.018)** |
| Macro-toxicity | Carcinogenicity | 0.712 (0.009) | 0.721 (0.034) | **0.732 (0.026)** | 0.726 (0.016) |
| | Hepatotoxicity | 0.778 (0.028) | **0.797 (0.022)** | 0.771 (0.015) | 0.780 (0.022) |
| | Rodent Acute Toxicity | 0.827 (0.123) | **0.894 (0.088)** | 0.817 (0.120) | 0.865 (0.169) |
| | Human Acute Toxicity | 0.659 (0.009) | **0.662 (0.011)** | 0.658 (0.009) | 0.661 (0.008) |
| Micro-toxicity | Ames mutagenicity | 0.820 (0.008) | **0.828 (0.010)** | 0.826 (0.005) | 0.825 (0.021) |
| | hERG inhibition | 0.760 (0.009) | 0.767 (0.004) | 0.778 (0.010) | **0.788 (0.014)** |
| | AR | 0.870 (0.017) | 0.877 (0.029) | 0.888 (0.025) | **0.905 (0.028)** |
| | ER | 0.817 (0.019) | 0.851 (0.023) | 0.822 (0.013) | **0.861 (0.016)** |
| | PPAR-γ | 0.802 (0.024) | 0.812 (0.016) | 0.815 (0.012) | **0.825 (0.013)** |
| | MMP | 0.864 (0.010) | 0.874 (0.009) | 0.884 (0.009) | **0.900 (0.005)** |
| | p53 | 0.789 (0.012) | 0.803 (0.009) | 0.804 (0.006) | **0.826 (0.004)** |
| | ARE | 0.794 (0.011) | 0.787 (0.007) | 0.814 (0.016) | **0.816 (0.003)** |
| | HSE | 0.787 (0.007) | 0.793 (0.011) | 0.807 (0.016) | **0.817 (0.011)** |
| | ATAD5 | 0.777 (0.013) | **0.827 (0.025)** | 0.819 (0.014) | 0.806 (0.009) |
| | AhR | 0.837 (0.009) | 0.848 (0.003) | 0.847 (0.010) | **0.850 (0.014)** |
| | Aromatase | 0.781 (0.015) | 0.786 (0.009) | 0.782 (0.008) | **0.789 (0.004)** |
| | Average | 0.797 | 0.812 | 0.807 | **0.819** |

[a] Without the integration of bioactivity datasets as auxiliary tasks.

# Comparison of the models in H-ADMET

Considering that ADMET datasets are usually small, on which deep learning models may not outperform traditional machine learning models, we designed the system with the functionality of selecting the best model from multiple models by their performance for the final prediction. We incorporated three different models in H-ADMET, namely LiteGEM, GINE+, and RF.

**Supplementary Table 8** and **Supplementary Table 9** summarise the performance of three incorporated models under random and scaffold splits. On classification tasks, LiteGEM achieves better average AUCs under both the random and scaffold split. LiteGEM outperforms RF by nearly 0.031 and 0.055 under the random and scaffold split, respectively. LiteGEM also performs significantly better than GINE+ under scaffold split. We can see that the advantage of LiteGEM is larger under scaffold split than under random split, demonstrating that the more challenging the dataset, the better LiteGEM performed compared to the baseline models. On regression tasks, LiteGEM also performs better than the other two models, although with a smaller number of endpoints, this advantage is not that convincing. The models we provide on the online platform is the best of the three models under random split, so that half of the endpoints in the online system are LiteGEM models, and about the other half are GINE+ models, while only two of them are the RF models (shown as bold in **Supplementary Table 8** and **Supplementary Table 9**). This further support that in the "big data" age, traditional machine learning models may not be the better choice for ADMET prediction tasks.

**Supplementary Table 8.** Performance of the three H-ADMET models on classification tasks [a].

| Category | ADMET Models | Random split | | | Scaffold split | | |
|---|---|---|---|---|---|---|---|
| | | RF | LiteGEM | GINE+ | RF | LiteGEM | GINE+ |
| Absorption | Caco-2 permeability | 0.853 (0.024) | **0.879 (0.009)** | 0.875 (0.053) | 0.794 (0.012) | **0.957 (0.013)** | 0.896 (0.023) |
| | P-glycoprotein substrate | 0.888 (0.045) | **0.891 (0.021)** | 0.852 (0.024) | - [b] | - | - |
| | P-glycoprotein inhibitor | 0.934 (0.009) | **0.947 (0.007)** | 0.937 (0.007) | 0.913 (0.003) | **0.932 (0.005)** | 0.916 (0.004) |
| | Oral bioavailability | 0.776 (0.044) | **0.803 (0.000)** | 0.731 (0.059) | 0.685 (0.012) | **0.752 (0.019)** | 0.701 (0.018) |
| Distribution | BBBP | 0.935 (0.032) | **0.944 (0.010)** | 0.938 (0.011) | 0.685 (0.010) | **0.699 (0.005)** | 0.695 (0.009) |
| Metabolism | CYP1A2 inhibitor | 0.924 (0.005) | **0.948 (0.003)** | 0.944 (0.009) | 0.871 (0.001) | **0.916 (0.003)** | 0.916 (0.004) |
| | CYP1A2 substrate | 0.896 (0.024) | 0.929 (0.017) | **0.949 (0.024)** | 0.741 (0.005) | **0.782 (0.013)** | 0.734 (0.023) |
| | CYP2C19 inhibitor | 0.890 (0.009) | **0.939 (0.005)** | 0.936 (0.005) | 0.818 (0.012) | 0.886 (0.004) | **0.893 (0.005)** |
| | CYP2C19 substrate | 0.872 (0.016) | 0.934 (0.025) | **0.945 (0.016)** | 0.644 (0.001) | **0.770 (0.021)** | 0.717 (0.001) |
| | CYP2C9 inhibitor | 0.881 (0.002) | 0.926 (0.006) | **0.934 (0.004)** | 0.812 (0.003) | **0.896 (0.004)** | 0.894 (0.002) |
| | CYP2C9 substrate | 0.881 (0.003) | **0.944 (0.009)** | 0.931 (0.009) | 0.688 (0.004) | **0.766 (0.017)** | 0.722 (0.010) |
| | CYP2D6 inhibitor | 0.875 (0.008) | **0.905 (0.010)** | 0.900 (0.008) | 0.846 (0.005) | **0.895 (0.004)** | 0.892 (0.005) |
| | CYP2D6 substrate | 0.906 (0.028) | 0.948 (0.011) | **0.956 (0.016)** | **0.779 (0.012)** | 0.765 (0.021) | 0.758 (0.019) |
| | CYP3A4 inhibitor | 0.901 (0.002) | **0.930 (0.003)** | 0.928 (0.007) | 0.869 (0.002) | 0.911 (0.008) | **0.911 (0.002)** |
| | CYP3A4 substrate | 0.937 (0.013) | 0.956 (0.015) | **0.967 (0.008)** | 0.718 (0.007) | **0.738 (0.028)** | 0.713 (0.036) |
| Excretion | Half life | 0.681 (0.031) | **0.736 (0.031)** | 0.715 (0.027) | 0.655 (0.012) | **0.741 (0.018)** | 0.723 (0.026) |
| Macro-toxicity | Carcinogenicity | 0.823 (0.016) | 0.815 (0.021) | **0.836 (0.042)** | 0.716 (0.012) | **0.726 (0.016)** | 0.654 (0.012) |
| | Hepatotoxicity | 0.766 (0.036) | **0.808 (0.005)** | 0.763 (0.008) | 0.741 (0.004) | **0.797 (0.022)** | 0.664 (0.016) |
| | Rodent Acute Toxicity | 0.696 (0.077) | 0.629 (0.196) | **0.717 (0.202)** | **0.718 (0.105)** | **0.808 (0.119)** | 0.698 (0.209) |
| | Human Acute Toxicity | 0.862 (0.009) | **0.873 (0.012)** | 0.863 (0.010) | 0.686 (0.007) | 0.662 (0.011) | **0.693 (0.006)** |
| Micro-toxicity | Ames mutagenicity | **0.909 (0.005)** | 0.900 (0.014) | 0.906 (0.006) | **0.844 (0.003)** | 0.828 (0.010) | 0.823 (0.029) |
| | hERG inhibition | 0.902 (0.006) | 0.907 (0.006) | **0.909 (0.006)** | 0.765 (0.003) | **0.788 (0.014)** | 0.782 (0.007) |
| | AR | 0.837 (0.040) | **0.858 (0.019)** | 0.832 (0.034) | 0.869 (0.011) | **0.905 (0.028)** | 0.860 (0.014) |
| | ER | 0.790 (0.064) | **0.848 (0.037)** | 0.844 (0.024) | 0.797 (0.009) | **0.861 (0.016)** | 0.836 (0.019) |
| | PPAR-γ | 0.834 (0.020) | 0.881 (0.046) | **0.889 (0.033)** | 0.736 (0.021) | **0.825 (0.013)** | 0.798 (0.019) |
| | MMP | 0.918 (0.008) | **0.951 (0.010)** | 0.946 (0.006) | 0.826 (0.006) | **0.900 (0.005)** | 0.881 (0.016) |
| | p53 | 0.866 (0.043) | **0.938 (0.016)** | 0.925 (0.017) | 0.761 (0.011) | **0.826 (0.004)** | 0.817 (0.009) |
| | ARE | 0.846 (0.012) | 0.884 (0.024) | **0.917 (0.017)** | 0.732 (0.009) | **0.816 (0.003)** | 0.794 (0.014) |
| | HSE | 0.814 (0.028) | **0.884 (0.028)** | 0.827 (0.036) | 0.759 (0.012) | 0.817 (0.011) | **0.819 (0.006)** |
| | ATAD5 | 0.844 | 0.921 | **0.939** | 0.732 | 0.827 | **0.844** |
| | | (0.023) | (0.019) | **(0.021)** | (0.015) | (0.025) | **(0.020)** |
| | AhR | 0.912 (0.010) | **0.922 (0.011)** | 0.909 (0.008) | 0.838 (0.008) | **0.850 (0.014)** | 0.846 (0.005) |
| | Aromatase | 0.829 (0.049) | 0.886 (0.036) | **0.913 (0.014)** | 0.682 (0.016) | **0.789 (0.004)** | 0.782 (0.016) |
| | Average | 0.859 | **0.890** | 0.887 | 0.765 | **0.820** | 0.796 |

[a] AUCs are listed for all three models and splitting methods.

[b] AUC for the endpoint of P-glycoprotein substrate under scaffold split cannot be calculated, since only one class remained in the test dataset.

**Supplementary Table 9.** Performance of the three H-ADMET models on regression tasks.

| Category | ADMET Models | Random split | | |
|---|---|---|---|---|
| | | RF | LiteGEM | GINE+ |
| Physicochemical property | Solubility | 0.804 (0.013) | **0.877 (0.010)** | 0.860 (0.007) |
| | pKa | 0.799 (0.018) | **0.847 (0.003)** | 0.799 (0.023) |
| Absorption | Human intestinal absorption | **0.786 (0.044)** | 0.754 (0.040) | 0.681 (0.063) |
| Distribution | PPB | 0.558 (0.006) | 0.736 (0.048) | **0.747 (0.025)** |
| | Average | 0.737 | **0.804** | 0.772 |

**Supplementary Table 10.** Contribution of each stage with the random split on classification tasks.

| Category | ADMET Models | Stage 3 | Stage 1+3 | Stage 2+3 | Stage 1+2+3 |
|---|---|---|---|---|---|
| Absorption | Caco-2 permeability | **0.879 (0.009)** | 0.851 (0.028) | 0.874 (0.008) | 0.875 (0.028) |
| | P-glycoprotein substrate | 0.786 (0.060) | 0.822 (0.016) | 0.860 (0.041) | **0.891 (0.021)** |
| | P-glycoprotein inhibitor | 0.926 (0.013) | 0.932 (0.005) | 0.940 (0.016) | **0.947 (0.007)** |
| | Oral bioavailability | 0.698 (0.001) | 0.698 (0.001) | 0.732 (0.000) | **0.803 (0.000)** |
| Distribution | BBBP | 0.926 (0.019) | 0.925 (0.015) | **0.944 (0.010)** | **0.944 (0.025)** |
| Metabolism | CYP1A2 inhibitor | 0.927 (0.005) | 0.934 (0.005) | **0.948 (0.003)** | **0.948 (0.006)** |
| | CYP1A2 substrate | 0.896 (0.018) | 0.901 (0.017) | 0.926 (0.013) | **0.929 (0.017)** |
| | CYP2C19 inhibitor | 0.899 (0.009) | 0.907 (0.006) | 0.938 (0.004) | **0.939 (0.005)** |
| | CYP2C19 substrate | 0.906 (0.022) | 0.912 (0.025) | **0.934 (0.025)** | 0.929 (0.016) |
| | CYP2C9 inhibitor | 0.888 (0.007) | 0.900 (0.006) | 0.923 (0.007) | **0.926 (0.006)** |
| | CYP2C9 substrate | 0.899 (0.032) | 0.898 (0.015) | **0.944 (0.009)** | 0.937 (0.015) |
| | CYP2D6 inhibitor | 0.876 (0.011) | 0.885 (0.007) | 0.904 (0.007) | **0.905 (0.010)** |
| | CYP2D6 substrate | 0.910 (0.021) | 0.909 (0.026) | **0.948 (0.011)** | 0.945 (0.009) |
| | CYP3A4 inhibitor | 0.902 (0.004) | 0.911 (0.003) | 0.929 (0.003) | **0.930 (0.003)** |
| | CYP3A4 substrate | 0.939 (0.019) | 0.941 (0.018) | **0.956 (0.015)** | 0.954 (0.013) |
| Excretion | Half life | 0.724 (0.037) | 0.719 (0.039) | 0.732 (0.031) | **0.736 (0.031)** |
| Macro-toxicity | Carcinogenicity | 0.778 (0.017) | 0.789 (0.015) | 0.807 (0.013) | **0.815 (0.021)** |
| | Hepatotoxicity | 0.781 (0.003) | 0.784 (0.022) | **0.808 (0.005)** | 0.795 (0.020) |
| | Rodent Acute Toxicity | 0.591 (0.103) | 0.610 (0.005) | 0.562 (0.184) | **0.629 (0.196)** |
| | Human Acute Toxicity | 0.829 (0.016) | 0.830 (0.015) | 0.857 (0.012) | **0.873 (0.012)** |
| Micro-toxicity | Ames mutagenicity | 0.890 (0.005) | 0.898 (0.013) | 0.898 (0.006) | **0.900 (0.014)** |
| | hERG inhibition | 0.885 (0.009) | 0.905 (0.007) | 0.906 (0.014) | **0.907 (0.006)** |
| | AR | 0.839 (0.056) | 0.845 (0.017) | **0.858 (0.019)** | 0.843 (0.025) |
| | ER | 0.788 (0.021) | 0.783 (0.035) | **0.848 (0.037)** | 0.843 (0.015) |
| | PPAR-γ | 0.815 (0.040) | 0.816 (0.046) | 0.861 (0.057) | **0.881 (0.046)** |
| | MMP | 0.932 (0.016) | 0.925 (0.009) | **0.951 (0.010)** | **0.951 (0.008)** |
| | p53 | 0.870 (0.041) | 0.883 (0.017) | 0.935 (0.019) | **0.938 (0.016)** |
| | ARE | 0.831 (0.023) | 0.835 (0.032) | **0.884 (0.024)** | 0.877 (0.021) |
| | HSE | 0.821 (0.037) | 0.823 (0.026) | **0.884 (0.028)** | 0.878 (0.028) |
| | ATAD5 | 0.814 (0.040) | 0.845 (0.030) | **0.921 (0.019)** | 0.911 (0.017) |
| | AhR | 0.904 (0.005) | 0.908 (0.020) | 0.919 (0.007) | **0.922 (0.011)** |
| | Aromatase | 0.847 (0.043) | 0.838 (0.033) | **0.886 (0.036)** | 0.882 (0.041) |
| | Average | 0.850 | 0.855 | 0.882 | **0.887** |

**Supplementary Table 11.** Contribution of each stage with the random split on regression tasks.

| Category | ADMET Models | Stage 3 | Stage 1+3 | Stage 2+3 | Stage 1+2+3 |
|---|---|---|---|---|---|
| Physicochemical property | Solubility | 0.812 (0.025) | 0.856 (0.020) | 0.851 (0.023) | **0.877 (0.010)** |
| | pKa | 0.835 (0.015) | 0.830 (0.012) | 0.799 (0.012) | **0.847 (0.003)** |
| Absorption | Human intestinal absorption | - [a] | 0.694 (0.054) | 0.725 (0.032) | **0.754 (0.040)** |
| Distribution | PPB | 0.610 (0.061) | 0.709 (0.029) | 0.709 (0.041) | **0.736 (0.048)** |
| | Average | 0.752 | 0.772 | 0.771 | **0.804** |

[a] The model performance was not reasonable when trained with only stage 3. $R^2$ of the best hyperparameter was less than 0.

# User manual

We incorporated our models into a user-friendly, web-based platform, on which ADMET prediction can be easily performed within two steps. The first step is model selection. Users may directly use our baseline model for their ADMET prediction, confirmed by making the selection from the drop-down menu (Supplementary Figure 3). This will allow the users to enter the second step, in which up to 1000 query compounds can be accepted as input (Supplementary Figure 4). Users may do so by drawing molecular structures, entering SMILES formula directly in the text box, or uploading a file containing all the SMILES strings in plain text, each line represents a molecule. Following task submission, users will be automatically taken to the result page once the prediction is completed. If they choose to leave the website or to run the program in background, the result page can always be accessible from the Task Management module. To provide an overall picture of all query molecules, compounds are arranged in a card view according to their input sequence, with each card displays 6 key endpoints of the compound (Supplementary Figure 5). This allows the users to do a quick scan on all results and locate molecules of interest efficiently. In addition, we offer an "Advanced Filters" panel on top of the page for the users to conveniently exclude undesired compounds before close examination. Upon clicking a card, the corresponding detail page exhibiting all ADMET endpoints will be expanded (Supplementary Figure 6). More explanation and interpretation for these endpoints can be found in the section of "Description of all endpoints" in this Supplementary and also on the website.

Alternatively, users may choose to build their own models by selecting "Train a new model" in the first step. This will expand extra entry fields as shown in Supplementary Figure 3. Users should choose to use either the pretrained regression or classification model depends on the type of dataset they provided. They will then be asked to upload the training data as a .csv file containing molecules in SMILES format, the labels of the endpoint and their units. A sample .csv file can be found on the website. Once the self-training process is completed, the newly generated model will be added to the drop-down menu in the first step. Users may repeatedly and conveniently use these models for their future predictions.

Altogether, H-ADMET offers an easy-to-use, highly flexible platform that generates comprehensive and accurate ADMET predictions.

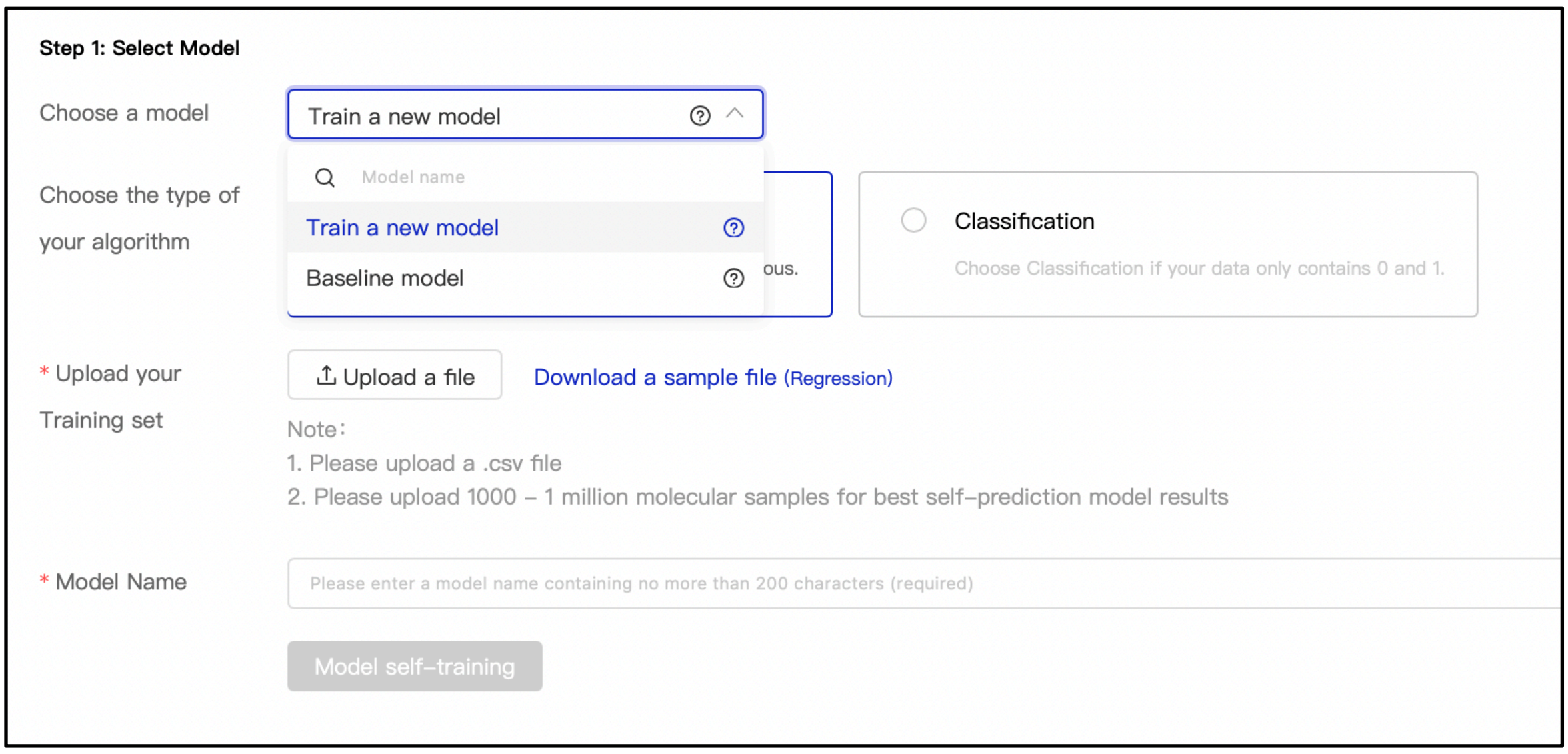

**Supplementary Figure 3.** The web interface of H-ADMET. Step 1: model selection.

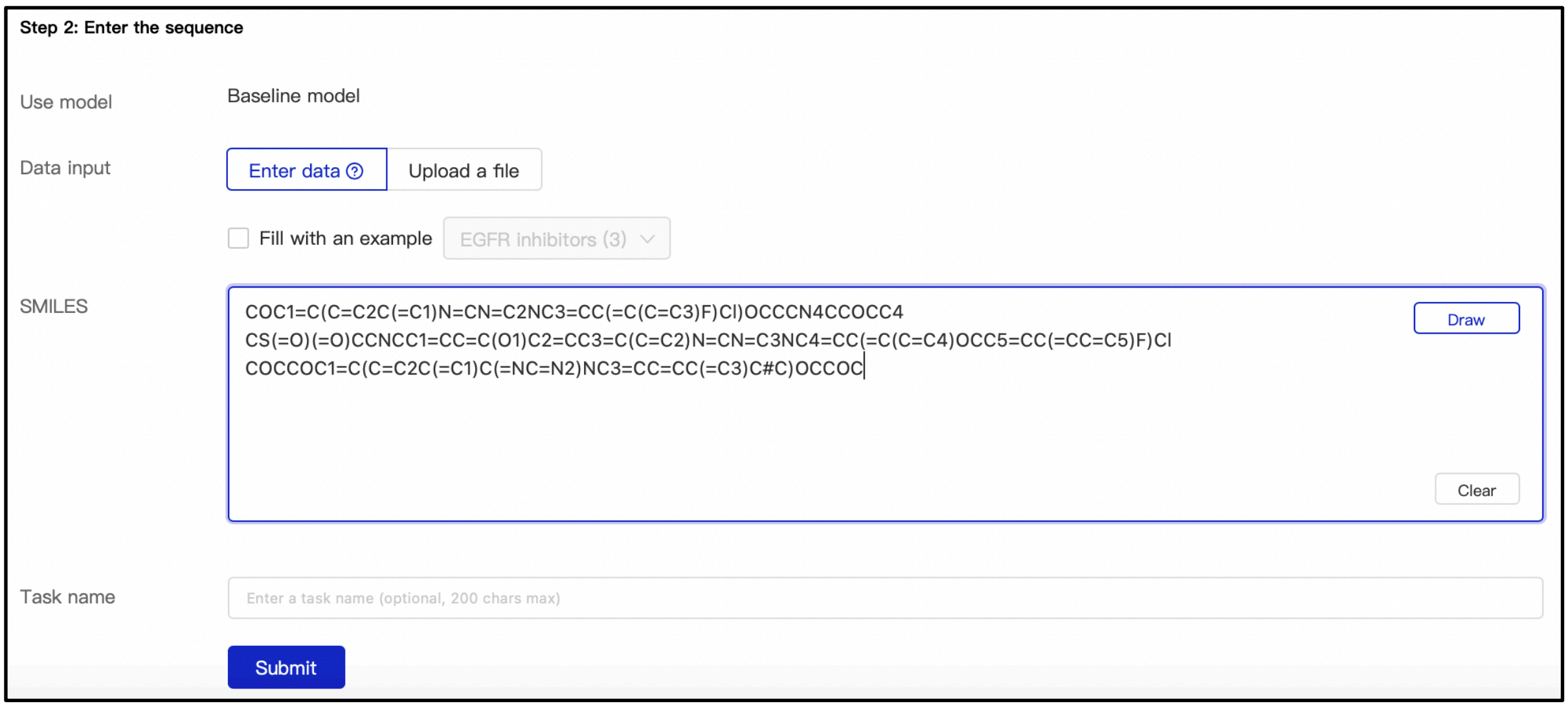

**Supplementary Figure 4.** The web interface of H-ADMET. Step 2: enter query compounds.

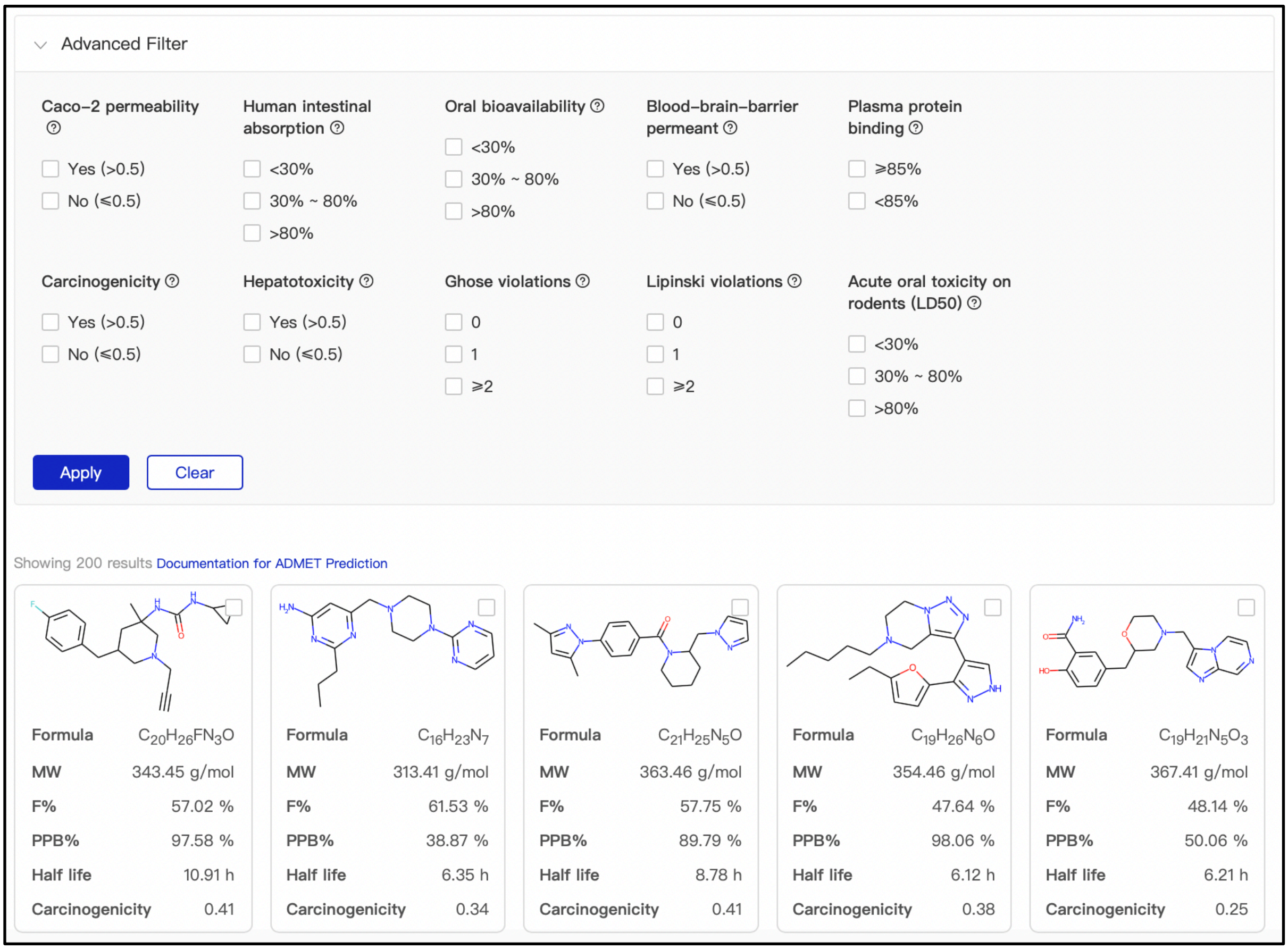

**Supplementary Figure 5.** The advanced filter panel and card view on the result page.

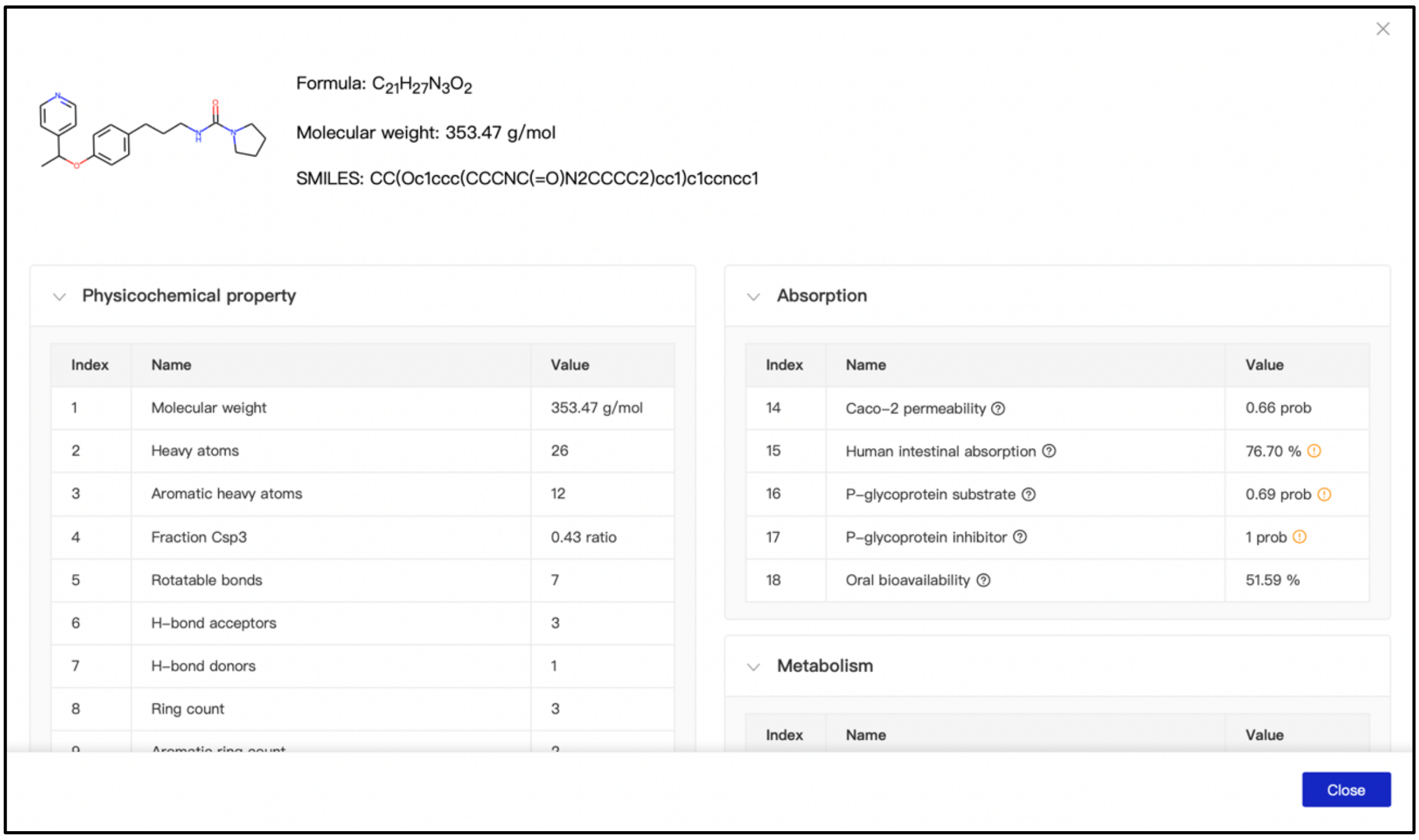

**Supplementary Figure 6.** A detailed page containing the full ADMET prediction results for a sample molecule.

# Supplementary references